\documentclass[showpacs,preprintnumbers,twocolumn,amsmath,amssymb]{revtex4}
\usepackage{graphicx}
\usepackage{dcolumn}
\usepackage{bm}
\usepackage{epsfig}
\usepackage[usenames]{color}

\newcommand{\dhd}{{\textstyle d}
\lower.03ex\hbox{\kern-0.40em$^{\scriptstyle-}$}\kern-0.08em{}}  

\newcommand{\half}{{1\over 2}}
\newcommand{\bu}{{\bullet}}

\newcommand{\bamma}{{\bar \gamma}}
\newcommand{\bark}{{\bar k}}

\newcommand{\baru}{{\bar u}}
\newcommand{\barv}{{\bar v}}

\newcommand{\barz}{{\bar z}}
\newcommand{\barP}{{\bar P}}

\newcommand{\cald}{{\cal D}}
\newcommand{\calf}{{\cal F}}   
\newcommand{\call}{{\cal L}}  
  
\newcommand{\calr}{{\cal R}}  
 
\newcommand{\calu}{{\cal U}} 
\newcommand{\calv}{{\cal V}} 
\newcommand{\calz}{{\cal Z}}

\newcommand{\halu}{\hat{\cal U}} 
\newcommand{\hsi}{\hat{\psi}} 
\newcommand{\hatj}{\hat{j}} 
\newcommand{\tildek}{\tilde{k}} 
 
\newcommand{\tildeP}{\tilde{P}} 
\newcommand{\tildez}{\tilde{z}} 
\begin{document}

\preprint{JLAB-THY-12-1596}
\preprint{NT-LBL-12-014}

\title{ Photon impact factor and $k_T$-factorization for DIS in the next-to-leading order}

\author{Ian Balitsky }
\affiliation{
Physics Dept., ODU, Norfolk VA 23529,  and\\
Theory Group, Jlab, 12000 Jefferson Ave, Newport News, VA 23606}
\email{balitsky@jlab.org}

\author{
Giovanni A. Chirilli}
\address{ Nuclear Science Division, Lawrence Berkeley National Laboratory, Berkeley, CA 94720, USA}

\date{\today}

\begin{abstract}

The photon impact factor for the BFKL pomeron is calculated in the next-to-leading order (NLO) 
approximation using the operator expansion in Wilson lines. The result is represented as a  
NLO $k_T$-factorization formula for the structure functions of small-$x$ deep inelastic scattering.

\end{abstract}



\maketitle

\section{\label{sec:in}Introduction }
It is well known that the small-$x$ behavior of structure functions of deep inelastic scattering is determined by the hard pomeron contribution. 
In the leading order the pomeron intercept is determined by the BFKL equation \cite{bfkl}  and the pomeron residue 
(the $\gamma^\ast\gamma^\ast$-pomeron vertex) 
is given by the so-called impact factor. To find the small-$x$ structure functions in the next-to-leading order, one needs to know both the pomeron intercept
and the impact factor. The NLO pomeron intercept was found many years  ago \cite{nlobfkl} but the analytic expression for the NLO impact factor is obtained
for the first time in the present paper.

 We calculate the NLO impact factor  using the high-energy operator expansion of T-product of two vector currents in Wilson lines (see e.g the reviews 
 \cite{mobzor,nlolecture}). Let us recall the general logic 
 of an operator expansion. In order to find a certain asymptotical behavior of an amplitude by OPE one should
  \begin{itemize}
\item Identify the relevant operators and  factorize an amplitude into a product of coefficient functions and matrix elements of these operators
\item Find the evolution equations of the operators with respect to the factorization scale
\item Solve these evolution equations
\item Convolute the solution  with the initial conditions for the evolution and get the amplitude.
\end{itemize}
Since we are interested in the small-$x$ asymptotics of deep inelastic scattering (DIS) 
it is natural to factorize in rapidity: we introduce the rapidity divide $\eta$ which separates the ``fast'' gluons 
from the ``slow'' ones. 

 As a first step, we integrate 
over gluons with rapidities $Y>\eta$ and leave the integration over $Y<\eta$ for the later time, see Fig. 1.
\begin{figure}[htb]
\includegraphics[width=33mm]{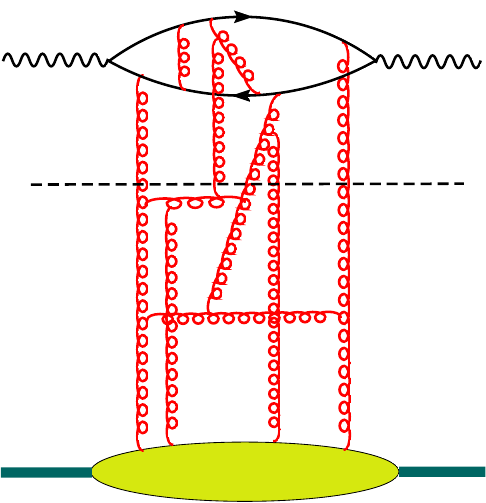}
\hspace{1cm}
\includegraphics[width=14mm]{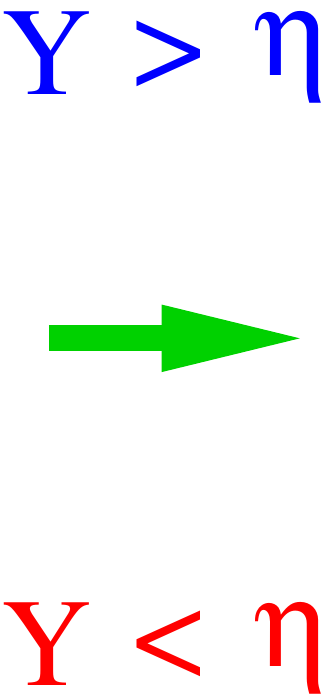}

\vspace{11mm}
\hspace{-0cm}
\includegraphics[width=60mm]{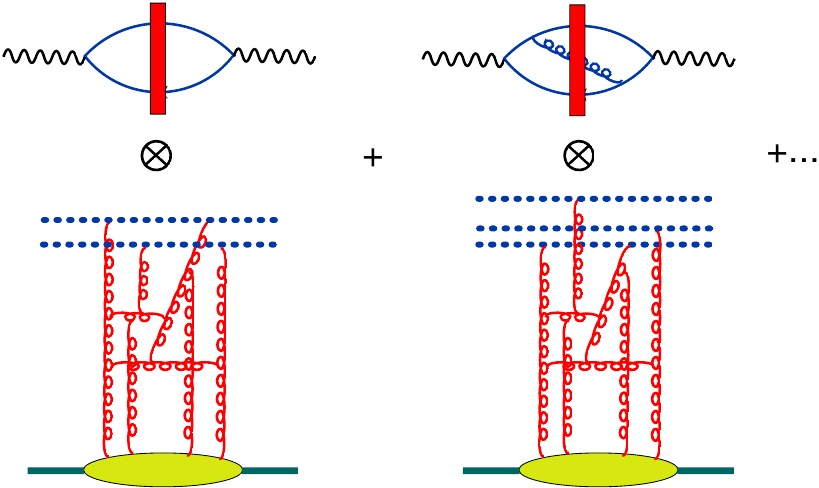}
\caption{Rapidity factorization. The impact factors with $Y>\eta$ are given by diagrams in the shock-wave background. 
Wilson-line operators with $Y<\eta$ are denoted by dotted lines.}
\label{fig:rapidityfac}
\end{figure}

\begin{figure}[htb]
\includegraphics[width=50mm]{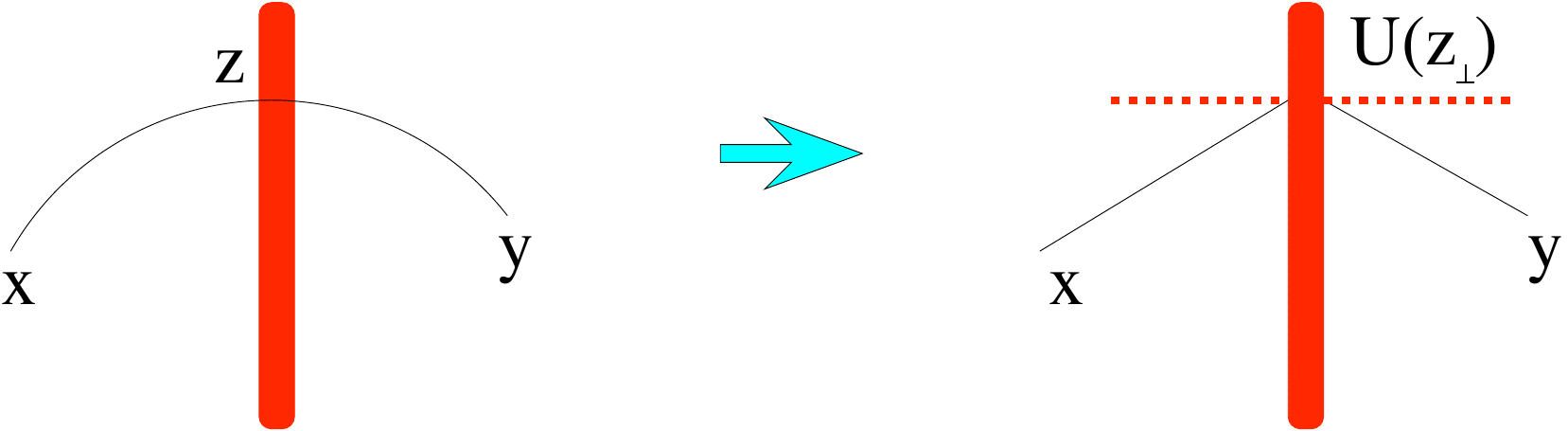}
\caption{Propagator in a shock-wave background}
\label{fig:shwaveprop}
\end{figure}

It is convenient to use the background field formalism: we integrate over gluons with $\alpha>\sigma=e^\eta$ and leave gluons with $\alpha<\sigma$ as a background field, to
be integrated over later.  Since the rapidities of the background
gluons are very different from the rapidities of gluons in our Feynman diagrams, the background field can be taken in the form of a shock wave due to the Lorentz contraction.
To derive the expression of a quark (or gluon) propagator in this shock-wave background we represent the propagator as a path integral over various trajectories,
each of them weighed with the gauge factor Pexp$(ig\int\! dx_\mu A^\mu)$ ordered along the propagation path. Now, since the shock wave is very thin, quarks (or gluons) do not
have time to deviate in transverse direction so their trajectory inside the shock wave can be approximated by a segment of the straight line. Moreover, since there is no external field 
outside the shock wave, the integral over the segment of straight line can be formally extended to $\pm\infty$ limits yielding the Wilson-line 
gauge factor
\begin{eqnarray}
&&\hspace{-0mm} 
 U^\eta_x~=~{\rm Pexp}\Big[ig\!\int_{-\infty}^\infty\!\! du ~p_1^\mu A^\sigma_\mu(up_1+x_\perp)\Big],
 \nonumber\\
 &&\hspace{-0mm} 
A^\eta_\mu(x)~=~\int\!d^4 k ~\theta(e^\eta-|\alpha_k|)e^{ik\cdot x} A_\mu(k)
\label{cutoff}
\end{eqnarray}
where  the  Sudakov variable $\alpha_k$ is defined as usual,  $k=\alpha_kp_1+\beta_kp_2+k_\perp$.
We define the light-like vectors $p_1$ and $p_2$ such that $q=p_1-x_Bp_2$  and $p=p_2+{m_N^2\over s}p_1$ where $q$ is the virtual photon 
momentum, $p$ is the momentum of the target particle, and $x_B=Q^2/s\ll 1$ is the Bjorken variable (at large energies $s\simeq 2p\cdot q$).
The structure of the propagator in a shock-wave background looks as follows (see Fig. \ref{fig:shwaveprop}): \\ 
$\big[$Free propagation from initial point $x$ to the point of intersection with the shock wave $z\big]$\\
$\times$ $\big[$Interaction
with the shock wave described by the Wilson-line operator $U_z\big]$\\
$\times$ $\big[$Free propagation from point of interaction $z$ to the final point $y\big]$. \\

The explicit form of quark propagator in a shock-wave background can be taken from Ref. \cite{npb96}
%
\begin{eqnarray}
&&\hspace{-1mm}
\langle T\{\hsi(x)\bar\hsi(y)\}\rangle_A~
\label{kvprop}\\
&&\hspace{-1mm}
\stackrel{x_\ast>0>y_\ast}{=}~-\!\int\! d^4z~\delta(z_\ast){(\not\! x-\not\! z)\over 2\pi^2(x-z)^4}\not\! p_2U_z{(\not\! z-\not\! y)\over 2\pi^2(x-z)^4}
\nonumber
\end{eqnarray}
As usual, we label operators by hats and $\langle\hat{\cal O}\rangle_A$ means the vacuum average of the operator $\hat{\cal O}$ in the presence of
an external field $A$.
Hereafter use the notations $x_\ast=p_2^\mu x_\mu=\sqrt{s\over 2}x^+$, $x_\bullet=p_1^\mu x_\mu=\sqrt{s\over 2}x^-$ (and our metric is (1,-1,-1,-1)).
Note that the Regge limit in the coordinate space can be achieved by rescaling
\begin{eqnarray}
&&\hspace{-1mm}
x~\rightarrow~\rho x_\ast {2\over s}p_1+x_\bu{2\over s\rho}p_2+x_\perp,
\nonumber\\
&&\hspace{-1mm}
 y~\rightarrow~\rho y_\ast {2\over s}p_1+y_\bu{2\over s\rho}p_2+y_\perp, ~~~~~~~
\label{reggelimita}
\end{eqnarray}
with $\rho\rightarrow \infty$,  see the discussion
in Refs. \cite{nlobfklconf, confamp}.

\begin{figure}[htb]
\includegraphics[width=40mm]{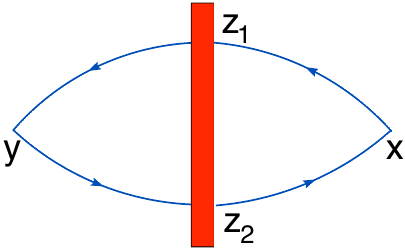}
\caption{Impact factor in the leading order. Solid lines represent quarks. \label{fig:loif}}
\end{figure}

The result of the integration over gluons with rapidities $Y>\eta$ gives the impact factor - the amplitude of the transition of virtual photon 
in two-Wilson-lines operators (sometimes called ``color dipole'').
The LO impact factor is a product of two propagators (\ref{kvprop}), see Fig. \ref{fig:loif} 
%
\begin{eqnarray}
&&\hspace{-1mm}
\langle T\{ \bar{\hsi}(x)\gamma^\mu \hsi(x)\bar{\hsi}(y)\gamma^\nu \hsi(y)\} \rangle_A~ =~
\label{loifa}\\
&&\hspace{-1mm}
=~{s^2\over 2^9\pi^6x_\ast^2y_\ast^2}\int d^2z_{1\perp}d^2z_{2\perp}
 {{\rm tr}\{U_{z_1}U^\dagger_{z_2}\}\over (\kappa\cdot\zeta_1)^3(\kappa\cdot\zeta_2)^3}
\nonumber\\
&&\hspace{-1mm}
\times~{\partial^2\over\partial x^\mu\partial y^\nu}
\big[2(\kappa\cdot\zeta_1)(\kappa\cdot\zeta_2)-\kappa^2(\zeta_1\cdot\zeta_2) \big]~+~O(\alpha_s)
\nonumber
\end{eqnarray}
Here we introduced the conformal vectors \cite{penecostalba,penedones} 
\begin{eqnarray}
&&\hspace{-5mm}
\kappa~=~\kappa_x-\kappa_y,~~~~~\kappa_x~=~{\sqrt{s}\over 2x_\ast}({p_1\over s}-x^2p_2+x_\perp)
\nonumber\\
&&\hspace{-1mm}
\zeta_i~=~\big({p_1\over s}+z_{i\perp}^2 p_2+z_{i\perp}\big),~\label{cratios}
\end{eqnarray}
and the notation $\calr~\equiv~{\kappa^2(\zeta_1\cdot\zeta_2)\over 2(\kappa\cdot\zeta_1)(\kappa\cdot\zeta_2)}$.
The above equation is explicitly M\"obius invariant. In addition, it is easy to check that ${\partial\over\partial x_\mu}$(r.h.s)=0.

Our goal is the NLO contribution to the r.h.s. of Eq. (\ref{loifa}), but first let us briefly discuss the three remaining steps of 
the high-energy OPE. 
The evolution equation for color dipoles has the form \cite{npb96,yura}
\begin{eqnarray}
&&\hspace{-2mm}
{d\over d\eta}{\rm tr}\{\hat{U}^\eta_{z_1} \hat{U}^{\dagger\eta}_{z_2}\}~
=~{\alpha_s\over 2\pi^2}
\!\int\!d^2z_3~
{z_{12}^2\over z_{13}^2z_{23}^2}
~[{\rm tr}\{\hat{U}^\eta_{z_1} \hat{U}^{\dagger\eta}_{z_3}\}
\label{nlobk}\\
&&\hspace{-2mm}
\times~{\rm tr}\{\hat{U}^\eta_{z_3} \hat{U}^{\dagger\eta}_{z_2}\}
-N_c{\rm tr}\{\hat{U}^\eta_{z_1} \hat{U}^{\dagger\eta}_{z_2}\}]   ~+~{\rm NLO~contribution}
\nonumber
\end{eqnarray}
(To save space, hereafter  $z_i$ stand for $z_{i\perp}$ so $z_{12}^2\equiv z_{12\perp}^2$ etc.)
The explicit form of the NLO contributions can be found in Refs. \cite{nlobk,nlobksym,nlolecture} while the agrument
of the coupling constant in the above equation (following from the NLO calculations) is discussed in Refs. (\cite{prd75,kw1}).

It is worth noting that we performed the OPE program outlined above for scattering of scalar ``particles'' in ${\cal N}=4$ SYM 
and obtained the explicit expression for the four-point correlator of scalar operators at high energies in the next-to-leading order \cite{confamp}.
In QCD the analytic solution of the evolution equation for color dipoles with running coupling constant is not known
at present. 
This prevents us from getting the explicit NLO amplitude as in ${\cal N}=4$ case.
We  can, however,  perform the first two steps in our OPE program discussed in the Introduction: calculate the coefficient function 
(impact factor) and find the evolution equation for color dipoles.  
The next two steps, solution of the evolution equation (\ref{nlobk}) with appropriate initial conditions and the eventual
comparison with experimental DIS data are discussed in many papers (see e.g. \cite{solutions}).
It is worth noting that, contrary to the evolution equation, the NLO correction to the impact factor
has nothing to do with running of the coupling constant - it starts at the NNLO level.
Thus, the argument of the coupling constant at the NLO level is determined
solely by the evolution equation for color dipoles.  For numerical estimates involving the impact factor  
one can take $\alpha_s(|x-y|)$ as the first approximation since the characteristic transverse distances in the impact factor are $\sim|x-y|$.

The paper is organized as follows: in Sect. 2 and 3 we calculate the NLO impact factor in the coordinate representation (the results of these Sections were
published previously in Brief Report \cite{bfrp}). The Mellin representation of the impact factor is presented in Sect. 4 and Sect. 5 contains the impact factor in
the momentum representation for the forward case corresponding to deep inelastic scattering. Finally, we present the NLO BFKL kernel
and discuss the $k_T$-factorization for DIS in Sect. 6.

\section{Calculation of the NLO impact factor}

Now we would like to repeat the steps of operator expansion discussed above to the NLO accuracy. 
A general form of the expansion of  T-product of the electromagnetic currents 
 in color dipoles looks as follows:
\begin{eqnarray}
&&\hspace{-1mm}
 (x-y)^4T\{\bar{\hsi}(x)\gamma^\mu \hsi(x)\bar{\hsi}(y)\gamma^\nu \hsi(y)\}~
 \nonumber\\
&&\hspace{-1mm}=~\int\! {d^2z_1d^2z_2\over z_{12}^4}~\Big\{I_{\mu\nu}^{\rm LO}(z_1,z_2)\big[1+{3\alpha_s\over 4\pi}c_F\big]
 {\rm tr}\{\hat{U}^\eta_{z_1}\hat{U}^{\dagger \eta}_{z_2}\}
 \nonumber\\
&&\hspace{-1mm}
+\int\! d^2z_3~I_{\mu\nu}^{\rm NLO}(z_1,z_2,z_3;\eta)
\nonumber\\
&&\hspace{-1mm}
\times~
[ {\rm tr}\{\hat{U}^\eta_{z_1}\hat{U}^{\dagger \eta}_{z_3}\}{\rm tr}\{\hat{U}^\eta_{z_3}\hat{U}^{\dagger \eta}_{z_2}\}
 -N_c {\rm tr}\{\hat{U}^\eta_{z_1}\hat{U}^{\dagger \eta}_{z_2}\}]\Big\}
 \label{opeq}
 \end{eqnarray}
For simplicity, we calculate at first the impact factor for one flavor of quarks with electric charge one and restore the trivial factor
$\sum e_i^2$ in Eq. (\ref{ktfacv})  below.

Unfortunately, in terms of Wilson-line approach there is no direct way to 
get the NLO impact factor for the BFKL pomeron. One needs first to find
the coefficient in front of the four-Wilson-line operator (which we will also call the NLO impact factor) 
 and then linearize it. 

The structure of the NLO contribution is clear from the topology of diagrams in the shock-wave background, see Fig. \ref{fig:nloif} below.
Also, the term $\sim~1+{3\alpha_s\over  4\pi}c_F$ can be restored from the requirement 
that at $U=1$ (no shock wave) one should get the perturbative series for the 
polarization operator $1+{3\alpha_s\over 4\pi}c_F+O(\alpha_s^2)$.

In our notations
\begin{eqnarray}
&&\hspace{-1mm}
I^{\rm LO}_{\mu\nu}(z_1,z_2)~=~
 {\calr^2\over \pi^6(\kappa\cdot\zeta_1)(\kappa\cdot\zeta_2)}
\nonumber\\
&&\hspace{-1mm}
\times~{\partial^2\over\partial x^\mu\partial y^\nu}
\big[(\kappa\cdot\zeta_1)(\kappa\cdot\zeta_2)-\half\kappa^2(\zeta_1\cdot\zeta_2) \big].~
\label{loif}
\end{eqnarray}
which corresponds to the well-known expression for the LO impact factor in the momentum space.

The NLO impact factor is given by the diagrams shown in Fig. \ref{fig:nloif}. 
\begin{figure}[htb]
\includegraphics[width=70mm]{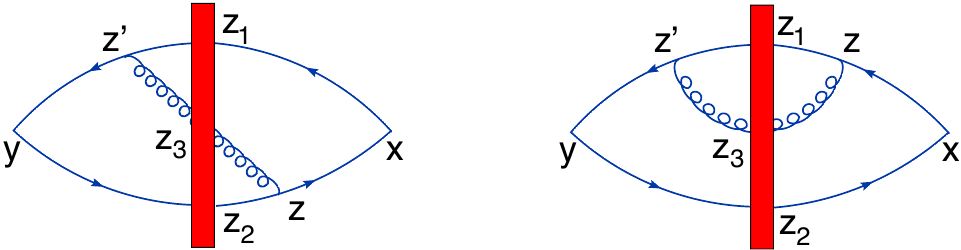}

\vspace{3mm}
\includegraphics[width=70mm]{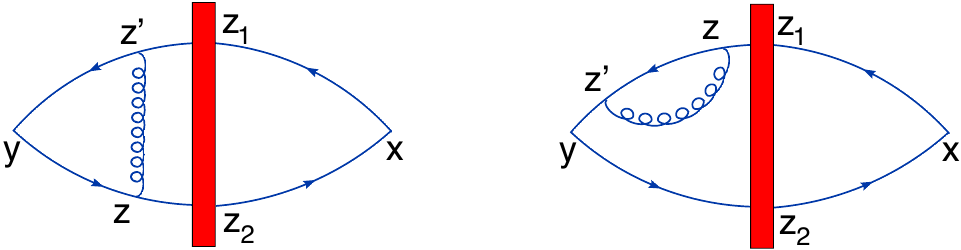}
\caption{Impact factor in the next-to-leading order. \label{fig:nloif}}
\end{figure}
The calculation of these diagrams
is similar to the calculation of the NLO impact factor for scalar currents in ${\cal N}=4$ SYM carried out in our previous paper \cite{nlobksym}.
The gluon propagator in the shock-wave background at $x_\ast>0>y_\ast$ in the light-like gauge $p_2^\mu A_\mu=0$ is given by \cite{prd99,balbel} 
\begin{eqnarray}
&&\hspace{-1mm}
\langle T\{ \hat{A}^a_\mu(x)\hat{A}^b_\nu(y)\}\rangle
~
\stackrel{x_\ast>0>y_\ast}{=}~-{i\over 2}\int d^4 z~\delta(z_\ast)~
\label{gluprop}\\
&&\hspace{-1mm}
\times~{x_\ast g^\perp_{\mu\xi}-p_{2\mu}(x-z)^\perp_\xi \over \pi^2[(x-z)^2+i\epsilon]^2}\;U^{ab}_{z_\perp}
{1\over\partial_\ast^{(z)}}~{y_*\delta^{\perp\xi}_\nu
 - p_{2\nu}(y-z)_\perp^\xi\over \pi^2[(z-y)^2+i\epsilon]^2}
 \nonumber
\end{eqnarray}
where ${1\over \partial_\ast}$ can be either ${1\over \partial_\ast+i\epsilon }$ or 
${1\over \partial_\ast-i\epsilon }$ which leads to the same result. (This is obvious for the leading order and 
correct in NLO after subtraction of the leading-order contribution, see Eq. (\ref{nloify}) below).

The diagrams in Fig. \ref{fig:nloif} can be calculated using the conformal integral
\begin{eqnarray}
&&\hspace{-3mm}
\int\! d^4z                                                             
~{\not\!{x}-\not\!{z}\over (x-z)^4}\gamma_\mu
{\not\!{z}-\not\!{y}\over(z-y)^4}{z_\nu\over z^4}-\mu\leftrightarrow\nu~
\nonumber\\
&&\hspace{-3mm} 
=~{\pi^2\over x^2y^2(x-y)^2}\Big[-\!\not\! x\gamma_\mu\!\not\! y\Big({x_\nu\over x^2}+{y_\nu\over y^2}\Big)
\nonumber\\
&&\hspace{-3mm} 
+~\half(\!\not\! x\gamma_\mu\gamma_\nu-\gamma_\mu\gamma_\nu\!\not\! y)
+2x_\mu y_\nu{\!\not\! x-\!\not\! y\over (x-y)^2}\Big]~-~\mu\leftrightarrow\nu~~~
\label{confintegral}
\end{eqnarray}
which gives the 3-point $\psi\Bar{\psi}F_{\mu\nu}$ Green function in the leading order in $g$. 
Using Eqs. (\ref{kvprop}), (\ref{gluprop}) and (\ref{confintegral}), performing integrals over $z_\bu$'s and 
taking traces one gets after some algebra the NLO contribution of diagrams in Fig. \ref{fig:nloif} 
in the form 
\begin{equation}
I^{\rm Fig.\ref{fig:nloif}}_{\mu\nu}(z_1,z_2,z_3)~=~\tilde{I}_1^{\mu\nu}(z_1,z_2,z_3)+I_2^{\mu\nu}(z_1,z_2,z_3)
\label{fig3vklad}
\end{equation}
where
\begin{eqnarray}
&&\hspace{-1mm}
~\tilde{I}_1^{\mu\nu}(z_1,z_2,z_3)~
\label{tildei1}\\
&&\hspace{-1mm}=~{\alpha_s\over 4\pi^2}I^{\rm LO}_{\mu\nu}(z_1,z_2){(\zeta_1\cdot\zeta_2)\over (\zeta_1\cdot\zeta_3) (\zeta_1\cdot\zeta_3)}
\!\int_0^\infty\!{d\alpha\over\alpha}e^{i\alpha {s\over 4}\sigma\calz_3}
\nonumber
\end{eqnarray}
and
\begin{eqnarray}
&&\hspace{-1mm}
(I_2)_{\mu\nu}(z_1,z_2,z_3)~
=~{\alpha_s\over 16\pi^8}{\calr^2\over(\kappa\cdot\zeta_1)(\kappa\cdot\zeta_2)}
\label{nloifoton}\\
&&\hspace{-1mm}
\times~
\Bigg\{{(\kappa\cdot\zeta_2)\over(\kappa\cdot\zeta_3)}
{\partial^2\over\partial x^\mu\partial y^\nu}
\Big[-{(\kappa\cdot\zeta_1)^2\over (\zeta_1\cdot\zeta_3)}+{(\kappa\cdot\zeta_1)(\kappa\cdot\zeta_2)\over (\zeta_2\cdot\zeta_3)}
\nonumber\\
&&\hspace{-1mm}
+~{(\kappa\cdot\zeta_1)(\kappa\cdot\zeta_3)(\zeta_1\cdot\zeta_2)\over  (\zeta_1\cdot\zeta_3) (\zeta_2\cdot\zeta_3)}
-{\kappa^2(\zeta_1\cdot\zeta_2)\over (\zeta_2\cdot\zeta_3)}\Big]+{(\kappa\cdot\zeta_2)^2\over (\kappa\cdot\zeta_3)^2}
\nonumber\\
&&\hspace{-1mm}
\times~{\partial^2\over\partial x^\mu\partial y^\nu}
\Big[{(\kappa\cdot\zeta_1)(\kappa\cdot\zeta_3)\over (\zeta_2\cdot\zeta_3)}-{\kappa^2(\zeta_1\cdot\zeta_3)\over 2(\zeta_2\cdot\zeta_3)}\Big]
+(\zeta_1\leftrightarrow\zeta_2)\Bigg\}
\nonumber
\end{eqnarray}
(recall that $z_{ij\perp}^2=2(\zeta_i\cdot\zeta_j)$ and $\calz_i={4\over\sqrt{s}}(\kappa\cdot\zeta_i)$). 
We obtained this expression at $x_\ast>0>y_\ast$ but  from the conformal structure of the result it is clear 
that this expression holds true at $x_\ast<0<y_\ast$ as well.

The integral over $\alpha$ in the r.h.s. of Eq. (\ref{tildei1}) diverges. This divergence reflects the fact that the 
contributions of the diagrams in Fig. \ref{fig:nloif} is not exactly the NLO impact factor since we must subtract the 
matrix element of the leading-order contribution. Indeed,  the NLO impact factor is a coefficient function defined according to Eq. (\ref{opeq}).
To find the NLO impact factor, we consider the operator equation (\ref{opeq}) in 
the shock-wave background (in the leading order $\langle\hat{U}_{z_3}\rangle_A=U_{z_3}$):
\begin{eqnarray}
&&\hspace{-1mm}
 \langle T\{\bar{\hsi}(x)\gamma^\mu \hsi(x)\bar{\hsi}(y)\gamma^\nu \hsi(y)\}\rangle_A~
  \nonumber\\
&&\hspace{-1mm}
-\int\! {d^2z_1d^2z_2\over z_{12}^4}~I_{\mu\nu}^{\rm LO}(x,y;z_1,z_2)
 \langle{\rm tr}\{\hat{U}^\eta_{z_1}\hat{U}^{\dagger \eta}_{z_2}\}\rangle_A
 \nonumber\\
&&\hspace{-1mm}
=~\int\! {d^2z_1d^2z_2\over z_{12}^4}d^2z_3~I_{\mu\nu}^{\rm NLO}(x,y;z_1,z_2,z_3;\eta)
 \nonumber\\
&&\hspace{-1mm}
[ {\rm tr}\{U_{z_1}U^\dagger_{z_3}\}{\rm tr}\{U_{z_3}U^\dagger_{z_2}\}
 -N_c {\rm tr}\{U_{z_1}U^\dagger_{z_2}\}]
 \label{opeq1}
 \end{eqnarray}
The NLO matrix element $ \langle T\{\bar{\hsi}(x)\gamma^\mu \hsi(x)\bar{\hsi}(y)\gamma^\nu \hsi(y)\}\rangle_A$ is given by Eq. (\ref{fig3vklad}) 
while the subtracted term is
\begin{eqnarray}
&&\hspace{-1mm}
{\alpha_s\over 2\pi^2}\!\int\! {d^2z_1d^2z_2\over z_{12}^4}~I_{\mu\nu}^{\rm LO}(z_1,z_2)\!\int_0^\sigma\!{d\alpha\over\alpha}\!\int\!d^2z_3~
{z_{12}^2\over z_{13}^2z_{23}^2}
\nonumber\\
&&\hspace{-2mm}
\times~
~[{\rm tr}\{U_{z_1} U^{\dagger}_{z_3}\}{\rm tr}\{ U_{z_3} U^{\dagger}_{z_2}\}
-~N_c{\rm tr}\{U_{z_1} U^{\dagger}_{z_2}\}]   
\label{subtracterm}
\end{eqnarray}
as follows from Eq. (\ref{nlobk}).
The $\alpha$ integration is cut from above by $\sigma=e^\eta$ in accordance with the definition of operators $\hat{U}^\eta$, see Eq.  (\ref{cutoff}). 
Subtracting  (\ref{subtracterm}) from  Eq. (\ref{fig3vklad}) we get
\begin{eqnarray}
&&\hspace{-3mm}
I^{\rm NLO}_{\mu\nu}(z_1,z_2,z_3;\eta)~=~I_1^{\mu\nu}(z_1,z_2,z_3;\eta)+I_2^{\mu\nu}(z_1,z_2,z_3),
\nonumber\\
&&\hspace{-3mm}
I_1^{\mu\nu}(x,y;z_1,z_2,z_3;\eta)~
\nonumber\\
&&\hspace{-3mm}
=~{\alpha_s\over 2\pi^2}I^{\rm LO}_{\mu\nu}(z_1,z_2){z_{12}^2\over z_{13}^2z_{23}^2}
\Big[\!\int_0^\infty\!{d\alpha\over\alpha}~e^{i\alpha {s\over 4}\calz_3}   
-\!\int_0^\sigma\!{d\alpha\over\alpha}\Big]
\nonumber\\
&&\hspace{-2mm}
=~-{\alpha_s\over 2\pi^2}I^{\rm LO}_{\mu\nu}
{z_{12}^2\over z_{13}^2z_{23}^2}
\big[\ln{\sigma s\over 4}\calz_3-{i\pi\over 2}+C\big]
 \label{nloify}
 \end{eqnarray}
where $C$ is the Euler constant.
Note that one should expect the NLO impact factor to be conformally invariant since it is determined by tree diagrams in Fig. \ref{fig:nloif}.
However, as discussed in Refs. \cite{nlobk,confamp,nlolecture}, formally the light-like Wilson lines are conformally (M\"obius) invariant but the 
longitudinal cutoff $\alpha<\sigma$ in Eq. (1) violates this property so the term $\sim\ln\sigma\calz_3$ in the r.h.s. of Eq. (\ref{nloify}) is not invariant. 
As was demonstrated in these papers, one can define a composite operator
in the form 
\begin{eqnarray}
&&\hspace{-2mm}
[{\rm tr}\{\hat{U}_{z_1}\hat{U}^{\dagger}_{z_2}\}\big]_a~
\label{confodipola}\\
&&\hspace{-2mm}
=~{\rm tr}\{\hat{U}^\sigma_{z_1}\hat{U}^{\dagger\sigma}_{z_2}\}
+~{\alpha_s\over 2\pi^2}\!\int\! d^2 z_3~{z_{12}^2\over z_{13}^2z_{23}^2}
[ {\rm tr}\{\hat{U}^\sigma_{z_1}\hat{U}^{\dagger\sigma}_{z_3}\}
\nonumber\\
&&\hspace{-2mm}
\times~{\rm tr}\{\hat{U}^\sigma_{z_3}\hat{U}^{\dagger\sigma}_{z_2}\}
-N_c {\rm tr}\{\hat{U}^\sigma_{z_1}\hat{U}^{\dagger\sigma}_{z_2}\}]
\ln {4az_{12}^2\over \sigma^2 sz_{13}^2z_{23}^2}~+~O(\alpha_s^2)
\nonumber
\end{eqnarray}
where $a$ is an arbitrary constant. It is convenient to choose the rapidity-dependent constant 
$a\rightarrow ae^{-2\eta}$ so that the 
$[{\rm tr}\{\hat{U}^\sigma_{z_1}\hat{U}^{\dagger\sigma}_{z_2}\}\big]_a^{\rm conf}$ 
does not depend on $\eta=\ln\sigma$ and all the rapidity dependence is encoded into $a$-dependence. Indeed,
it is easy to see that ${d\over d\eta}[{\rm tr}\{\hat{U}_{z_1}\hat{U}^{\dagger}_{z_2}\}\big]_a^{\rm conf}~=~0$ 
and ${d\over da}[{\rm tr}\{\hat{U}_{z_1}\hat{U}^{\dagger}_{z_2}\}\big]_a^{\rm conf}$ is determined by the
NLO BK kernel which is a sum of the conformal part and the running-coupling part
with our $O(\alpha_s^2)$ accuracy\cite{nlobksym,nlolecture}. 

Rewritten in terms of composite  dipoles (\ref{confodipola}), the operator expansion (\ref{opeq})  takes the form:
\begin{eqnarray}
&&\hspace{-1mm}
 T\{\bar{\hsi}(x)\gamma^\mu \hsi(x)\bar{\hsi}(y)\gamma^\nu \hsi(y)\}~
\nonumber\\
&&\hspace{-1mm}=~\int\! {d^2z_1d^2z_2\over z_{12}^4}~\Big\{I_{\mu\nu}^{\rm LO}(z_1,z_2)\big[1+{3\alpha_s\over 4\pi}c_F\big]
[ {\rm tr}\{\hat{U}_{z_1}\hat{U}^{\dagger }_{z_2}\}]_a
 \nonumber\\
&&\hspace{-1mm}
+\int\! d^2z_3~I_{\mu\nu}^{\rm NLO}(z_1,z_2,z_3;a)
\nonumber\\
&&\hspace{-1mm}
\times~
[ {\rm tr}\{\hat{U}_{z_1}\hat{U}^{\dagger}_{z_3}\}{\rm tr}\{\hat{U}_{z_3}\hat{U}^{\dagger}_{z_2}\}
 -N_c {\rm tr}\{\hat{U}_{z_1}\hat{U}^{\dagger}_{z_2}\}]_a\Big\}
 \label{opeconf}
 \end{eqnarray}
We need to choose the  ``new rapidity cutoff'' $a$ in such a way that all the energy dependence is included in the matrix element(s) of 
Wilson-line operators so the impact factor should not depend on energy. A suitable
choice of $a$ is given by $a_0=-\kappa^{-2}+i\epsilon=-{4x_\ast y_\ast\over s(x-y)^2}+i\epsilon$ so we obtain
\begin{eqnarray}
&&\hspace{-1mm}
 (x-y)^4T\{\bar{\hsi}(x)\gamma^\mu \hsi(x)\bar{\hsi}(y)\gamma^\nu \hsi(y)\}~
 \label{opeconfin}\\
&&\hspace{-1mm}
=\int\! {d^2z_1d^2z_2\over z_{12}^4}~\Big\{I^{\mu\nu}_{\rm LO}(z_1,z_2)\big[1+{3\alpha_s\over 4\pi}c_F\big]
[ {\rm tr}\{\hat{U}_{z_1}\hat{U}^{\dagger }_{z_2}\}]_{a_0}
 \nonumber\\
&&\hspace{-1mm}
+\int\! d^2z_3\Big[ {\alpha_s\over 4\pi^2}{z_{12}^2\over z_{13}^2z_{23}^2}
\Big(\ln{\kappa^2(\zeta_1\cdot\zeta_3)(\zeta_1\cdot\zeta_3)\over 2(\kappa\cdot\zeta_3)^2(\zeta_1\cdot\zeta_2)}
-2C\Big)I^{\mu\nu}_{\rm LO}
\nonumber\\
&&\hspace{-1mm}
+~I_2^{\mu\nu}\Big]
[ {\rm tr}\{\hat{U}_{z_1}\hat{U}^{\dagger}_{z_3}\}{\rm tr}\{\hat{U}_{z_3}\hat{U}^{\dagger}_{z_2}\}
 -N_c {\rm tr}\{\hat{U}_{z_1}\hat{U}^{\dagger}_{z_2}\}]_{a_0}\Big\}
 \nonumber
 \end{eqnarray}
Here the composite dipole $[{\rm tr}\{\hat{U}^\sigma_{z_1}\hat{U}^{\dagger\sigma}_{z_2}\}]_{a_0}$ is given by Eq. (\ref{confodipola}) with
$a_0=-{4 x_\ast y_\ast\over s(x-y)^2}+i\epsilon$ while  $I^{\mu\nu}_{\rm LO}(z_1,z_2)$ and $I_2^{\mu\nu}(z_1,z_2,z_3)$ are given by Eqs. (\ref{loif})
and (\ref{nloifoton}), respectively.

\section{NLO impact factor for the BFKL pomeron}

For the studies of DIS with the linear NLO BFKL equation (up to two-gluon accuracy)  we need the linearized version of Eq. (\ref{opeconfin}).
If we define 
\begin{equation}
\halu_{a}(z_1,z_2)=1-{1\over N_c}[{\rm tr}\{\hat{U}_{z_1}\hat{U}^{\dagger}_{z_2}\}]_a
\label{kalu}
\end{equation}
and consider the linearization
\begin{eqnarray}
&&\hspace{-0mm}
{1\over N_c^2} {\rm tr}\{\hat{U}_{z_1}\hat{U}^{\dagger}_{z_3}\}{\rm tr}\{\hat{U}_{z_3}\hat{U}^{\dagger}_{z_2}\}
 -{1\over N_c} {\rm tr}\{\hat{U}_{z_1}\hat{U}^{\dagger}_{z_2}\}]_{a_0}~
 \nonumber\\
&&\hspace{-0mm}
\simeq~\halu(z_1,z_2)-\halu(z_1,z_3)-\halu(z_2,z_3)
 \nonumber
 \end{eqnarray}
one of the integrals over $z_i$ in the r.h.s. of Eq. (\ref{opeconfin}) can be performed. 
The result is
\begin{eqnarray}
&&\hspace{-4mm}
{1\over N_c}(x-y)^4T\{\bar{\hsi}(x)\gamma^\mu \hsi(x)\bar{\hsi}(y)\gamma^\nu \hsi(y)\}~
\label{ifresult}\\
&&\hspace{-4mm}
=~{\partial\kappa^\alpha\over\partial x^\mu}{\partial\kappa^\beta\over\partial y^\nu}
\!\int\! {dz_1 dz_2\over z_{12}^4}~\halu_{a_0}(z_1,z_2)\big[{\cal I}_{\alpha\beta}^{\rm LO}\big(1+{\alpha_s\over\pi}\big)+ {\cal I}_{\alpha\beta}^{\rm NLO}\big]
\nonumber
\end{eqnarray}
where
\begin{equation}
{\cal I}^{\alpha\beta}_{\rm LO}(x,y;z_1,z_2)~=~ \calr^2{g^{\alpha\beta}(\zeta_1\cdot\zeta_2)-\zeta_1^\alpha\zeta_2^\beta-\zeta_2^\alpha\zeta_1^\beta\over \pi^6(\kappa\cdot\zeta_1)(\kappa\cdot\zeta_2)}
\label{loif1}
\end{equation}
(see Eq. (\ref{loif})) and 
\begin{eqnarray}
&&\hspace{-1mm}
{\cal I}^{\alpha\beta}_{\rm NLO}(x,y;z_1,z_2)~=~
{\alpha_sN_c\over 4\pi^7}\calr^2
\Bigg\{
{\zeta_1^\alpha\zeta_2^\beta+\zeta_1\leftrightarrow \zeta_2\over (\kappa\cdot\zeta_1)(\kappa\cdot\zeta_2)}
\nonumber\\
&&\hspace{-1mm}
\times~
\Big[4{\rm Li}_2(1-{\cal R})-{2\pi^2\over 3}+{2\ln {\cal R}\over 1-{\cal R}}+{\ln {\cal R}\over {\cal R}}-4\ln {\cal R} +{1\over 2{\cal R}}
\nonumber\\
&&\hspace{-1mm}
-~2+2(\ln {1\over  {\cal R}}+{1\over {\cal R}}-2)\big(\ln {1\over {\cal R}}+2C\big)-4C-{2C\over {\cal R}}\Big]
\nonumber\\
&&\hspace{-1mm}
+\Big({\zeta_1^\alpha\zeta_1^\beta\over(\kappa\cdot\zeta_1)^2}
+\zeta_1\leftrightarrow\zeta_2\Big)
\Big[{\ln {\cal R}\over {\cal R}}-{2C\over {\cal R}}+2{\ln {\cal R}\over 1-{\cal R}}-{1\over 2{\cal R}}\Big]
\nonumber\\
&&\hspace{-1mm}
+~\Big[{\zeta_1^\alpha\kappa^\beta+\zeta_1^\beta\kappa^\alpha\over (\kappa\cdot\zeta_1)\kappa^2}
+\zeta_1\leftrightarrow \zeta_2\Big]
\Big[-2{\ln {\cal R}\over 1-{\cal R}}-{\ln {\cal R}\over {\cal R}}
\nonumber\\
&&\hspace{-1mm}
+\ln {\cal R}-{3\over 2{\cal R}}+{5\over 2}+2C+{2C\over {\cal R}}\Big] 
-{2\over\kappa^2}\Big(g^{\alpha\beta}-2{\kappa^\alpha\kappa^\beta\over\kappa^2}\Big)
\nonumber\\
&&\hspace{-1mm}
+~{g^{\alpha\beta}(\zeta_1\cdot\zeta_2)\over(\kappa\cdot\zeta_1)(\kappa\cdot\zeta_2)}\Big[{2\pi^2\over 3}-4{\rm Li}_2(1-{\cal R})
-2\big(\ln {1\over  {\cal R}}+{1\over {\cal R}}
\nonumber\\
&&\hspace{-1mm}
+~{1\over 2{\cal R}^2}-3\big)\big(\ln {1\over {\cal R}}+2C\big)
+6\ln {\cal R}-{2\over  {\cal R}}+2 +{3\over 2{\cal R}^2}\Big]\Bigg\}
\nonumber\\
\label{nloifresult}
\end{eqnarray}
where Li$_2(z)$ is the dilogarithm.  Here one easily recognizes five conformal tensor structures discussed in Ref. \cite{penecostalba2}.

While it is easy to see that 
\begin{eqnarray}
&&\hspace{-1mm}
{d\over dx_\mu}{1\over (x-y)^4}{\partial\kappa^\alpha\over\partial x^\mu}{\partial\kappa^\beta\over\partial y^\nu}
{\cal I}_{\alpha\beta}^{\rm LO}(x,y;z_i)~=~0
\end{eqnarray}
one should be careful when checking the electromagnetic gauge invariance in the next-to-leading order. The reason is that
the  composite dipole $\halu^{a_0}(z_1,z_2)$ depends on $x$ via the rapidity cutoff $a_0=-{4x_\ast y_\ast\over s(x-y)^2}$ so 
from Eq. (\ref{ifresult}) we get
\begin{eqnarray}
&&\hspace{-4mm}
{\partial\over \partial x_\mu}{1\over (x-y)^4}{\partial\kappa^\alpha\over\partial x^\mu}{\partial\kappa^\beta\over\partial y^\nu}
\!\int\! {dz_1 dz_2\over z_{12}^4}~\halu_{a_0}(z_1,z_2){\cal I}_{\alpha\beta}^{\rm NLO}(x,y;z_i)
\nonumber\\
&&\hspace{-4mm}
=~\Big(2{(x-y)^\mu\over (x-y)^2}-{p_2^\mu\over x_\ast}\Big){1\over (x-y)^4}
\nonumber\\
&&\hspace{-4mm}
\times~{\partial\kappa^\alpha\over\partial x^\mu}{\partial\kappa^\beta\over\partial y^\nu}\!\int\! {dz_1 dz_2\over z_{12}^4}
{\cal I}^{\rm LO}_{\alpha\beta}(x,y;z_i)
\left.a{d\over da}\halu_a(z_1,z_2)\right|_{a_0}
\end{eqnarray}
Using the leading-order BFKL equation in the dipole form (linearization of Eq. (\ref{nlobk}))
\begin{eqnarray}
&&\hspace{-1mm}
a{d\over da}\hat{\cal U}_a(z_1,z_2)
~=~{\alpha_sN_c\over 4\pi^2}\!\int\!d^2z_3{z_{12}^2\over z_{13}^2z_{23}^2}
\nonumber\\
&&\hspace{-1mm}
\times~
\big[\hat{\cal U}_a(z_1,z_3)+\hat{\cal U}_a(z_2,z_3)-\hat{\cal U}_a(z_1,z_2)\big]
\label{nlolin}
\end{eqnarray}
we obtain the following consequence of gauge invariance
\begin{eqnarray}
&&\hspace{-1mm}
{\partial\over \partial x_\mu}{1\over (x-y)^4}{\partial\kappa^\alpha\over\partial x^\mu}{\partial\kappa^\beta\over\partial y^\nu}
{\cal I}_{\alpha\beta}^{\rm NLO}(x,y;z_i)
\label{checkgaugeinv}\\
&&\hspace{-1mm}
=~{\alpha_s\over \pi^8}
{ y_\ast\over x_\ast (x-y)^6}\calr^3
\Big[\big({1\over 2\calr}-3-\ln \calr\big){\partial\ln\kappa^2\over \partial y^\nu}
\nonumber\\
&&\hspace{-1mm}
+\Big({\ln \calr\over\calr}+{5\over 2\calr}-{1\over 2\calr^2}\Big){\partial\over \partial y^\nu}[\ln(\kappa\cdot\zeta_1)+\ln(\kappa\cdot\zeta_2)]\Big]
\nonumber
\end{eqnarray}
We have verified that the expression (\ref{nloifresult}) satisfies the above equation.

\section{Photon impact factor in the Mellin representation}

In preparation for Fourier transformation we calculated the Mellin transform of the photon impact factor (\ref{nloifresult}). 
We project the impact factor on the conformal eigenfunctions of the BFKL equation \cite{lip86}
\begin{equation}
\hspace{-0mm}
E_{\nu,n}(z_{10},z_{20})~
=~\Big[{\tilde{z}_{12}\over \tilde{z}_{10}\tilde{z}_{20}}\Big]^{\half+i\nu+{n\over 2}}
\Big[{\barz_{12}\over \barz_{10}\barz_{20}}\Big]^{\half+i\nu-{n\over 2}}
\label{eigenfunctions}
\end{equation}
(here $\tilde{z}=z_x+iz_y,\barz=z_x-iz_y$, $z_{10}\equiv z_1-z_0$ etc.). Since electromagnetic currents  are vectors,  
the only non-vanishing contribution comes from projection on the eigenfunctions with spin $0$ and spin 2. The spin-0 projection
has the form (throughout the paper we reserve the notation $\gamma$ for $\half+i\nu$):\\
\begin{eqnarray}
&&\hspace{-2mm}
\big(1+{3\alpha_s\over 4\pi}c_F\big){\cal J}_{\alpha\beta}^{\rm LO}(x,y;z_0,\nu)
+{\cal J}_{\alpha\beta}^{\rm NLO}(x,y;z_0,\nu)
\nonumber\\
&&\hspace{-2mm}
=\int\!{d^2z_1d^2z_2\over z_{12}^4}{\partial\kappa^\lambda\over\partial x^\alpha}{\partial\kappa^\rho\over\partial y^\beta}
\big[{\cal I}_{\lambda\rho}^{\rm LO}(x,y;z_1,z_2)
\nonumber\\
&&\hspace{-2mm}
\times~\big(1+{3\alpha_s\over 4\pi}c_F\big)+~{\cal I}_{\lambda\rho}^{\rm NLO}(x,y;z_1,z_2)\big]
\Big({z_{12}^2\over z_{10}^2z_{20}^2}\Big)^\gamma
\nonumber\\
&&\hspace{-2mm}
=~{1\over 4\pi^4}B(\bamma,\bamma)\Gamma(\gamma+1)\Gamma(2-\gamma)
\Big({\kappa^2\over(2\kappa\cdot\zeta_0)^2}\Big)^\gamma
\nonumber\\
&&\hspace{-2mm}
\times~
\Big\{-{\gamma\bamma \over 3}(2S_1+S_2)_{\mu\nu}
\Big[1+{3\alpha_s\over 4\pi}c_F+{\alpha_s N_c\over 2\pi}F_1(\gamma)\Big]
\nonumber\\
&&\hspace{-2mm}
-~2S_{2\mu\nu}\Big[1+{3\alpha_s\over 4\pi}c_F+{\alpha_s N_c\over 2\pi}F_2(\gamma)\Big]
\nonumber\\
&&\hspace{-2mm}
+~2\gamma(S_2-S_3)_{\mu\nu}\Big[1+{3\alpha_s\over 4\pi}c_F+{\alpha_s N_c\over 2\pi}F_3(\gamma)\Big]
\nonumber\\
&&\hspace{-2mm}
+~{\bamma\gamma^2\over (3-2\gamma)}\big(-{1\over 3}S_1-{2\over 3}S_2
\nonumber\\
&&\hspace{-2mm}
+~S_3-2S_4\big)_{\mu\nu}
\Big[1+{3\alpha_s\over 4\pi}c_F+{\alpha_s N_c\over 2\pi}F_4(\gamma)\Big]
\label{cormel}\\
&&\hspace{-2mm}
+~(S_1+S_2)_{\mu\nu}(2+\bamma\gamma)\Big[1+{3\alpha_s\over 4\pi}c_F+{\alpha_s N_c\over 2\pi}F_5(\gamma)\Big]\Big\}
\nonumber
\end{eqnarray}
where $\gamma\equiv\half +i\nu$, $\bamma\equiv 1-\gamma=\half -i\nu$  and 
\begin{eqnarray}
&&\hspace{-1mm}
F_1(\gamma)~=~F(\gamma)+{\chi_\gamma\over\gamma\bamma},
\nonumber\\
&&\hspace{-1mm}
F_2(\gamma)~=~F(\gamma)-1
+{1\over 2\gamma\bamma}+\chi_\gamma,
\nonumber\\
&&\hspace{-1mm}
F_3(\gamma)~=~F(\gamma)+\half\chi_\gamma,
\nonumber\\
&&\hspace{-1mm}
F_4(\gamma)~=~F(\gamma)-{6\over \gamma\bamma}
+{3\over\gamma^2\bamma^2}-{2\chi_\gamma\over\gamma\bamma},
\nonumber\\
&&\hspace{-1mm}
F_5(\gamma)~=~F(\gamma)
+{3\bamma\gamma\chi_\gamma+1-2\bamma\gamma\over \gamma\bamma(2+\bamma\gamma)},
\nonumber\\
&&\hspace{-1mm}
F(\gamma)~=~{2\pi^2\over 3}+1-{2\pi^2\over\sin^2\pi\gamma}-2C\chi_\gamma+{\chi_\gamma-2\over\bamma\gamma}
\label{cormelf}
\end{eqnarray}
and 
\begin{eqnarray}
&&\hspace{-1mm}
S_1^{\mu\nu}~\equiv~{\partial^2\ln\kappa^2\over\partial x_\mu\partial y_\nu},
~~~~~~~S_2^{\mu\nu}~\equiv~{\partial\ln\kappa^2\over\partial x_\mu}{\partial\ln\kappa^2\over\partial y_\nu},
\nonumber\\
&&\hspace{-1mm}
S_3^{\mu\nu}~\equiv~{\partial\ln\kappa^2\over\partial x_\mu}{\partial\ln\kappa\cdot\zeta_0\over\partial y_\nu}+
{\partial\ln\kappa\cdot\zeta_0\over\partial x_\mu}{\partial\ln\kappa^2\over\partial y_\nu},
\nonumber\\
&&\hspace{-1mm}
S_4^{\mu\nu}~\equiv~{\partial\ln\kappa\cdot\zeta_0\over\partial x_\mu}{\partial\ln\kappa\cdot\zeta_0\over\partial y_\nu}.
\label{esses}
\end{eqnarray}

The contribution of spin 2 in the t-channel has the form\\
\begin{eqnarray}
&&\hspace{-0mm}
\big(1+{3\alpha_s\over 4\pi}c_F\big){\cal J}_{2,\alpha\beta}^{\rm LO}(x,y;z_0,\nu)
+{\cal J}_{2,\alpha\beta}^{\rm NLO}(x,y;z_0,\nu)
\nonumber\\
&&\hspace{-0mm}
=~\int\!{d^2z_1d^2z_2\over z_{12}^4}{\partial\kappa^\alpha\over\partial x^\mu}{\partial\kappa^\beta\over\partial y^\nu}
\big[\big(1+{3\alpha_s\over 4\pi}c_F\big){\cal I}_{\alpha\beta}^{\rm LO}(x,y;z_1,z_2)
\nonumber\\
&&\hspace{10mm}
+~{\cal I}_{\alpha\beta}^{\rm NLO}(x,y;z_1,z_2)\big]
\Big({z_{12}^2\over z_{10}^2z_{20}^2}\Big)^\gamma{\tildez_{12}\over\tildez_{10}\tildez_{20}}{\barz_{10}\barz_{20}\over\barz_{12}}
\nonumber\\
&&\hspace{-0mm}
=~-{1\over 2\pi^4(x-y)^2}B(2-\gamma,2-\gamma)\Gamma(\gamma+2)\Gamma(3-\gamma)
\nonumber\\
&&\hspace{20mm}
\times~\Big[1+{3\alpha_s\over 4\pi}c_F+{\alpha_s N_c\over 2\pi}F_6(\gamma)\Big] S_5^{\mu\nu}
\label{cormelspin2}
\end{eqnarray}
where 
\begin{equation}
\hspace{-0mm}
F_6(\gamma)~=~F(\gamma)+{2C\over \bamma\gamma}-{2\over\bamma\gamma}+{2\over\bamma^2\gamma^2}
+3{1+\chi_\gamma-{1\over\gamma\bamma}\over 2+\bamma\gamma}-{\chi_\gamma\over\bamma\gamma}
\end{equation}
and
\begin{eqnarray}
&&\hspace{-0mm}
S_5^{\mu\nu}~
\equiv~\Big[g^{\mu 1}-ig^{\mu 2}-2(x-z_0)^\mu{\partial\over\partial\tildez_0}\ln\kappa\cdot\zeta_0
\nonumber\\
&&\hspace{10mm}
+~{4p_2^\mu\over\sqrt{s}}{(\kappa_x\cdot\zeta_0)(\kappa_y\cdot\zeta_0)\over (\kappa\cdot\zeta_0)}
{\partial\over\partial\tildez_0}\ln{\kappa_x\cdot\zeta_0\over\kappa_y\cdot\zeta_0}\Big]
\nonumber\\
&&\hspace{10mm}
\times~
\Big[g^{\nu 1}-ig^{\nu 2}-2(y-z_0)^\nu{\partial\over\partial\tildez_0}\ln\kappa\cdot\zeta_0
\nonumber\\
&&\hspace{10mm}
+~{4p_2^\nu\over\sqrt{s}}{(\kappa_x\cdot\zeta_0)(\kappa_y\cdot\zeta_0)\over (\kappa\cdot\zeta_0)}
{\partial\over\partial\tildez_0}\ln{\kappa_x\cdot\zeta_0\over\kappa_y\cdot\zeta_0}\Big]
\end{eqnarray}

Using the decomposition of the product of the transverse $\delta$-functions in conformal 3-point functions (\ref{eigenfunctions})
\begin{eqnarray}
&&\hspace{-1mm}      
\delta^{(2)}(z_1-z_3)\delta^{(2)}(z_2-z_4)~
\label{lobzor120}\\
&&\hspace{-1mm}
=\!\sum_{n=-\infty}^\infty\!\int\! {d\nu\over \pi^4}~{\nu^2+{n^2\over 4}\over z_{12}^2z_{34}^2}
\int\! d^2z_0~ E_{\nu,n}(z_{10},z_{20})E^\ast_{\nu,n}(z_{30},z_{40})
\nonumber
 \end{eqnarray}
we obtain
\begin{eqnarray}
&&\hspace{-1mm}      
\halu(z_1,z_2)~=~\!\int\! {d\nu\over \pi^2}\!\int\!d^2z_0\Big({z_{12}^2\over z_{10}^2z_{20}^2}\Big)^\gamma
\Big\{\nu^2\halu(z_0,\nu)
\nonumber\\
&&\hspace{-1mm}
+~(\nu^2+1)
\Big[{\tildez_{12}\barz_{10}\barz_{20}\over\tildez_{10}\tildez_{20}\barz_{12}}\hat{\bar{\calu}}^{(2)}(z_0,\nu)
+{\barz_{12}\tildez_{10}\tildez_{20}\over\barz_{10}\barz_{20}\tildez_{12}}\hat{\tilde{\calu}}^{(2)}(z_0,\nu)\Big]\Big\}
\label{decompu}
\nonumber\\
 \end{eqnarray}
where
\begin{eqnarray}
&&\hspace{-1mm}
 \hat{\cal U}_a(\nu,z_0)~\equiv~\!\int\! {d^2z_1d^2z_2\over \pi^2z_{12}^4}
\Big({z_{12}^2\over z_{10}^2z_{20}^2}\Big)^{\bamma}~ \hat{\cal U}_{a_0}(z_1,z_2)
 \nonumber\\
&&\hspace{-1mm}
 {\hat{\bar{\cal U}}}^{(2)}_{a}(\nu,z_0)~\equiv~\!\int\! {d^2z_1d^2z_2\over \pi^2z_{12}^4}
\Big({z_{12}^2\over z_{10}^2z_{20}^2}\Big)^{-\gamma}\!{\barz_{12}^2\over\barz_{10}^2\barz_{20}^2} \hat{\cal U}_{a}(z_1,z_2)
\nonumber\\
&&\hspace{-1mm}
 {\hat{\tilde{\cal U}}}^{(2)}_{a}(\nu,z_0)~\equiv~\!\int\! {d^2z_1d^2z_2\over \pi^2z_{12}^4}
\Big({z_{12}^2\over z_{10}^2z_{20}^2}\Big)^{-\gamma}\!{\tildez_{12}^2\over\tildez_{10}^2\tildez_{20}^2} \hat{\cal U}_{a}(z_1,z_2)
\nonumber\\
\label{Us}
\end{eqnarray}
is a composite dipole (\ref{confodipola}) in the Mellin representation.

 Substituting the decomposition (\ref{lobzor120})) in Eq. (\ref{ifresult}) we get the high-energy OPE in the form
\begin{eqnarray}
&&\hspace{-1mm}      
{1\over N_c} (x-y)^4T\{\hatj^\mu(x)\hatj^\nu(y)\}~
\label{opemellin}\\
&&\hspace{-1mm}
=~\int\! {d\nu\over\pi^2}\!\int\! d^2z_0~\Big\{\nu^2
\Big[\big(1+{3\alpha_s\over 4\pi}c_F\big){\cal J}_{\alpha\beta}^{\rm LO}(x,y;z_0,\nu)
\nonumber\\
&&\hspace{21mm}
+{\cal J}_{\alpha\beta}^{\rm NLO}(x,y;z_0,\nu)\Big]
 \hat{\cal U}_{a_0}(\nu,z_0)
 \nonumber\\
&&\hspace{-1mm}
+(\nu^2+1)\Big[\big\{\big(1+{3\alpha_s\over 4\pi}c_F\big){\cal J}_{2,\alpha\beta}^{\rm LO}(x,y;z_0,\nu)
\nonumber\\
&&\hspace{21mm}
+{\cal J}_{2,\alpha\beta}^{\rm NLO}(x,y;z_0,\nu)\big\}{\hat{\bar{\cal U}}}_{a_0}(\nu,z_0)+{\rm c.c.}\Big] \Big\}
\nonumber
 \end{eqnarray}
 The Eq. (\ref{opemellin}) and its Fourier transform (\ref{opemoment}) are the main results of this paper.

At this point it is instructive to check again the photon gauge invariance ${\partial\over\partial x^\mu}T\{\hatj^\mu(x)\hatj^\nu(y)\}~=~0$.
Since $a_0=-\kappa^2+i\epsilon$ we need to differentiate $\hat{\cal U}_{a_0}(\nu,z_0)$ too:
\begin{equation}
{\partial\over\partial x^\mu}\hat{\cal U}_{a_0}(\nu,z_0)~=~\left.a{d\over da}\hat{\cal U}_{a}(\nu,z_0)\right|_{a=a_0}
{\partial\over\partial x^\mu}\ln a_0
\label{fla37}
\end{equation}

Let us start with spin-2 contribution. It is easy to see that 
$$
{\partial\over\partial x^\mu}{S_5^{\mu\nu}\over(x-y)^6}~=~0,~~~~~~~
S_5^{\mu\nu}{\partial\over\partial x^\mu}\ln a_0 ~=~0
$$
and therefore the second term in r.h.s. of Eq. (\ref{opemellin}) is gauge invariant (recall that $a_0=-\kappa^{-2}+i\epsilon$).

For spin-0 part we need to use eq. (\ref{fla37}).  
 Since $\hat{\cal U}^a(\nu,z_0)$ are the projections of color dipoles on the eigenfunctions (\ref{eigenfunctions}) of the BFKL equation,
 the evolution equation (\ref{nlolin}) simplifies to
\begin{equation}
2a{d\over da} \hat{\cal U}_a(\nu,z_0)~=~\omega(\nu)\hat{\cal U}^a(\nu,z_0)
\label{dipevolution}
\end{equation}
where $\omega(\nu)~=~{\alpha_s N_c\over\pi}\chi_\gamma$ is the BFKL pomeron intercept  (as usual $\gamma=\half+i\nu$).
We obtain
\begin{equation}
{\partial\over\partial x^\mu}\hat{\cal U}_{a_0}(\nu,z_0)~=~-{\omega(\nu)\over 2}\hat{\cal U}_{a_0}(\nu,z_0)
{\partial\over\partial x^\mu}\ln\kappa^2
\label{dmucalu}
\end{equation}
In the leading order the derivative (\ref{dmucalu}) does not contribute so the formula for gauge invariance is simply
$$
{\partial\over\partial x^\mu}{\cal J}_{\alpha\beta}^{\rm LO}(x,y;z_0,\nu)~=~0
$$ 
It is easy to demonstrate that ${\cal J}_{\alpha\beta}^{\rm LO}$ in the r.h.s. of Eq. (\ref{cormel}) satisfies this requirement.

In the NLO we need both ${\cal J}_{\alpha\beta}^{\rm NLO}$ and $\omega {\cal J}_{\alpha\beta}^{\rm LO}$ parts so the
requirement for electromagnetic gauge invariance takes the form
\begin{equation}
{\partial\over\partial x^\mu}{\cal J}_{\alpha\beta}^{\rm NLO}(x,y;z_0,\nu)
~=~{\omega(\nu)\over 2}{\cal J}_{\alpha\beta}^{\rm LO}(x,y;z_0,\nu)
{\partial\ln\kappa^2\over\partial x^\mu}
\label{gaugeinvmellin}
\end{equation}
We have checked that the r.h.s. of Eq. (\ref{cormel})  satisfies this equation.

\section{Photon impact factor in the momentum space}

 In general, the rapidity evolution of color dipoles is non-linear but in this paper we assume that we can linearize it to
 the dipole form of the BFKL equation, like in the case of scattering of two virtual photons. 
 Moreover,  we will consider only the forward case which corresponds to deep inelastic scattering. 
 In this case, one may write down the high-energy OPE in the form of $k_T$-factorization formula
\begin{eqnarray}
&&\hspace{-0mm}
\int\! d^4x ~e^{iqx} \langle p|T\{\hatj_{\mu}(x)\hatj_\nu(0)\}|p\rangle~
\nonumber\\
&&\hspace{-0mm}
=~{s\over 2}\int\! {d^2k_\perp \over 4\pi^2}I_{\mu\nu}(q,k_\perp) \langle\!\langle p|\halu(k_\perp)|p\rangle\!\rangle
\label{ktfac}
\end{eqnarray}
where 
$$
\langle\halu(k_\perp)\rangle~=~\int\! d^2 x ~ e^{-i(k,x)_\perp}\langle\halu(x_\perp,0)\rangle,
$$
$q=p_1+{q^2\over s}p_2$ and  $p=p_2+{m^2\over s}p_1$ is the target's momentum.
 The reduced matrix element $ \langle\!\langle p|\halu(k)|p\rangle\!\rangle$
 is defined as 
\begin{eqnarray}
&&\hspace{-0mm}
\langle p|\halu(k)|p+\beta p_2\rangle~=~2\pi\delta(\beta)\langle\!\langle p|\halu(k)|p\rangle\!\rangle
\nonumber\\
&&\hspace{-0mm}
 \langle\!\langle p|\halu(k)|p\rangle\!\rangle~=~\int\!d^2z~e^{-i(k,z)_\perp} \langle\!\langle p|\halu(z,0)|p\rangle\!\rangle
 \label{defumom}
\end{eqnarray}
where the factor $2\pi\delta(\beta)$ reflects the fact that the forward matrix element of the operator $U_xU^\dagger_y$ contains an 
unrestricted integration along the $p_1$. Our goal in this Section is to find the impact factor $I_{\mu\nu}(q,k_\perp)$
 in the next-to-leading order.

Since our ``energy scale'' $a_0=-\kappa^{-2}$ for color dipoles depends on $x$ and $y$ ,
to perform the Fourier transformation of the OPE (\ref{opemellin}) one should express $\hat{\cal U}_{a_0}$ in terms of $\hat{\cal U}_{a_m}$ with
$a_m$ independent of coordinates $x$ and $y$. A suitable choice is $a_m={1/ x_B}$. With this choice, the impact factor does not scale with $s$ and
all the energy dependence is included in martix elements of color dipoles. This is similar to the choice $\mu^2=Q^2$ for the DGLAP evolution:
the coefficient functions in front of the light-ray operators 
 will not depend on $Q^2$ (except for $\alpha_s(Q^2)$ of course) and all the $Q^2$ dependence is shifted to parton densities.
The leading-order evolution of a color dipole $\hat{\cal U}_{a}$ 
is given by Eq. (\ref{dipevolution})
\begin{eqnarray}
&&\hspace{-1mm}
\hat{\cal U}_{a_0}(\nu,z_0)~=~\hat{\cal U}_{a_m}(\nu,z_0)\big({a_0x_B}\big)^{\omega(\nu)\over 2}, ~~
\nonumber\\
&&\hspace{-1mm}
\hat{\cal U}^{(2)}_{a_0}(\nu,z_0)~=~\hat{\cal U}^{(2)}_{a_m}(\nu,z_0)\big({a_0x_B}\big)^{\omega(2,\nu)\over 2},
\end{eqnarray}
so the Fourier transform of Eq. (\ref{opemellin}) yields\\
\begin{eqnarray}
&&\hspace{-2mm}
{1\over N_c}\!\int d^4 x d^4y ~\delta(y_\bu)~e^{iq\cdot (x-y)}~T\{\bar\psi\gamma_\mu\psi(x)\bar\psi\gamma_\nu\psi(y)\}
\label{opemoment}\\
&&\hspace{-2mm}
=~\!\int\! {d\nu\over \pi^3}\!\int\!d^2z_0~
\Big\{{\Gamma\big(\bamma+{\omega(\nu)\over 2}\big)
\Gamma^2\big(2-\gamma+{\omega(\nu)\over 2}\big)\Gamma(2-\gamma)
\over\Gamma(4-2\gamma+\omega(\nu))\Gamma\big(2+\gamma+{\omega(\nu)\over 2}\big)}
\nonumber\\
&&\hspace{-2mm}
\times~
{2\gamma-1\over 2\gamma+1}
\Big[(\gamma\bamma+2)P_1^{\mu\nu}\Big(1+{3\alpha_s\over 4\pi}c_F+{\alpha_sN_c\over 2\pi}\Phi_1(\nu)\Big)
\nonumber\\
&&\hspace{-2mm}
+(3\gamma\bamma+2)P_2^{\mu\nu}\Big(1+{3\alpha_s\over 4\pi}c_F+{\alpha_sN_c\over 2\pi}\Phi_2(\nu)\Big)
\Big]\halu_{a_m}(z_0,\nu)
\nonumber\\
&&\hspace{-2mm}
-~{\bamma\Gamma(3-\gamma)\Gamma\big(\bamma+{\omega(2,\nu)\over 2}\big)
\Gamma^2\big(2-\gamma+{\omega(2,\nu)\over 2}\big)
\over 2\Gamma(4-2\gamma+\omega(2,\nu))\Gamma\big(2+\gamma+{\omega(2,\nu)\over 2}\big)}
\nonumber\\
&&\hspace{-2mm}
\times~\Big[ {\hat{\bar{\cal U}}}_{a_m}^{(2)}(\nu,z_0)
\bar{P}^{\mu\nu}
+~{\hat{\tilde{\cal U}}}_{a_m}^{(2)}(\nu,z_0)\tildeP^{\mu\nu}\Big]
\nonumber\\
&&\hspace{-2mm}
\times~\Big(1+{3\alpha_s\over 4\pi}c_F+{\alpha_sN_c\over 2\pi}F_6(\nu)\Big)\Big\}{\Gamma^2(\bamma)\over\Gamma(2\bamma)}
{(Q^2)^{\gamma-1}\Gamma(2+\gamma)\over  4^{\gamma+1}}
\nonumber
\end{eqnarray}
where
\begin{eqnarray}
&&\hspace{-1mm}
P_1^{\mu\nu}~=~g^{\mu\nu}-{q_\mu q_\nu\over q^2}
\label{Pes}\\
&&\hspace{-1mm}
P_2^{\mu\nu}~=~{1\over q^2}\Big(q^\mu-{p_2^\mu q^2\over q\cdot p_2}\Big)\Big(q^\nu-{p_2^\nu q^2\over q\cdot p_2}\Big)
\nonumber\\
&&\hspace{-1mm}
\barP^{\mu\nu}~=~\big(g^{\mu 1}-ig^{\mu 2}\big)\big(g^{\nu 1}-ig^{\nu 2}\big)
\nonumber\\
&&\hspace{-1mm}
\tildeP^{\mu\nu}~=~\big(g^{\mu 1}+ig^{\mu 2}\big)\big(g^{\nu 1}+ig^{\nu 2}\big)
\nonumber
\end{eqnarray}
and \\
\begin{eqnarray}
&&\hspace{-1mm}
\Phi_1(\nu)~=~F(\gamma)+{3\chi_\gamma\over 2+\bamma\gamma}-{25\over 18(2-\gamma)}
\nonumber\\
&&\hspace{11mm}
+~{1\over 2\bamma}-{1\over 2\gamma}-{7\over 18(1+\gamma)}
+{10\over 3(1+\gamma)^2}
\nonumber\\
&&\hspace{-1mm}
\Phi_2(\nu)~=~F(\gamma)+{1\over 2\bamma\gamma}
\nonumber\\
&&\hspace{11mm}
-~{7\over 2(2+3\bamma\gamma)}
+{\chi_\gamma\over 1+\gamma}+{\chi_\gamma(1+3\gamma)\over 2+3\bamma\gamma}
\label{Phis}
\end{eqnarray}

The last step is to take forward matrix element and rewrite Eq. (\ref{opemoment}) in the $k_T$-factorized form (\ref{ktfac}). Using
\begin{eqnarray}
&&\hspace{-0mm}
\int\! d^2z_0\langle\hat{\calu}(z_0,\nu)\rangle~
\nonumber\\
&&\hspace{-0mm}
=~{\Gamma(-\bamma)\Gamma(1-2\gamma)\Gamma^2(\gamma)
\over 4^\bamma\Gamma(2-\gamma)\Gamma^2(\bamma)\Gamma(2\gamma)}
\!\int\! \dhd^2k ~k^{2\bamma}\langle\halu(k)\rangle,
\nonumber\\
&&\hspace{-0mm}
\int\! d^2z_0\langle{\hat{\tilde{\calu}}}^{(2)}(z_0,\nu)\rangle~=~-{4^{\gamma-1}\Gamma(\gamma)\Gamma(3-2\gamma)\Gamma^2(1+\gamma)
\over \Gamma(3-\gamma)\Gamma^2(2-\gamma)\Gamma(1+2\gamma)}
\nonumber\\
&&\hspace{21mm}
\times~
\!\int\! \dhd^2k ~{\bark\over\tildek}k^{2\bamma}\langle{\hat{\tilde{\calu}}}^{(2)}(k)\rangle
\label{flaxz}
\end{eqnarray}
and canceling the trivial factor $\int\! dz_\ast =2\pi\delta(0)$ on both sides we get
\begin{eqnarray}
&&\hspace{-0mm}
\!\int d^4 x ~e^{iqx}~\langle p|T\{\hat{j}_\mu(x+{2\over s}z_\ast p_1)\hat{j}_\nu({2\over s}z_\ast p_1)\}|p\rangle
\nonumber\\
&&\hspace{-0mm}
=~N_c{s\over 2}\!\int\! {d\nu\over 16\pi^3}\!\int\!{d^2k\over 4\pi^2}~\big({k^2\over Q^2}\big)^{\bamma}
{\Gamma^3(\bamma)\Gamma^3(\gamma)\over\Gamma(2\bamma)\Gamma(2\gamma)}
{\langle\!\langle p|\halu(k)|p\rangle\!\rangle\over 3+4\bamma\gamma}
\nonumber\\
&&\hspace{-0mm}
\times~\Big\{{\Gamma\big(\bamma+{\omega(\nu)\over 2}\big)
\Gamma^2\big(2-\gamma+{\omega(\nu)\over 2}\big)\Gamma(4-2\gamma)\Gamma(2+\gamma)
\over
\Gamma(\bamma)\Gamma^2(2-\gamma)\Gamma(4-2\gamma+\omega(\nu))\Gamma\big(2+\gamma+{\omega(\nu)\over 2}\big)}
\nonumber\\
&&\hspace{0mm}
\times~
\Big[(\gamma\bamma+2)P_1^{\mu\nu}\Big(1+{3\alpha_s\over 4\pi}c_F+{\alpha_sN_c\over 2\pi}\Phi_1(\nu)\Big)
\nonumber\\
&&\hspace{10mm}
+~(3\gamma\bamma+2)P_2^{\mu\nu}\Big(1+{3\alpha_s\over 4\pi}c_F+{\alpha_sN_c\over 2\pi}\Phi_2(\nu)\Big)\Big]
\nonumber\\
&&\hspace{-0mm}
+~{\Gamma\big(\bamma+{\omega(2,\nu)\over 2}\big)
\Gamma^2\big(2-\gamma+{\omega(2,\nu)\over 2}\big)\Gamma(4-2\gamma)\Gamma(2+\gamma)
\over 
\Gamma(\bamma)\Gamma^2(2-\gamma)\Gamma(4-2\gamma+\omega(2,\nu))\Gamma\big(2+\gamma+{\omega(2,\nu)\over 2}\big)}
\nonumber\\
&&\hspace{10mm}
\times~{\bamma\gamma\over 2}P_3^{\mu\nu}\Big(1+{3\alpha_s\over 4\pi}c_F+{\alpha_sN_c\over 2\pi}F_6(\nu)\Big)\Big\}
\label{opemomenta}
\end{eqnarray}
where
$$
P_3^{\mu\nu}~=~{1\over k_\perp^2}\big[ \tildek^2\bar{P}^{\mu\nu}
+~\bark^2\tildeP^{\mu\nu}\big]~=~4{k_\perp^\mu k_\perp^\nu\over k_\perp^2}+2g_\perp^{\mu\nu} 
$$
There is a subtle point in in the Fourier transform of Eq. (\ref{opemoment}): the contribution of infinite $z_0$ doubles the result from finite $z_0$ as shown in the Appendix.

Note that $\omega(\nu)~=~{\alpha_sN_c\over\pi}\chi_\gamma$ and $\omega(2,\nu)~=~{\alpha_sN_c\over\pi}\big(\chi_\gamma-{1\over\bamma\gamma}\big)$ are of order $\alpha_s$ so one should expand the expressions like $\Gamma(4-2\gamma+\omega(\nu))$ 
in Eq. (\ref{opemoment}) up to the first order in $\omega(\nu)$.
Using definitions (\ref{ktfac}) and  (\ref{defumom})  one obtains the impact factor in the form
\begin{eqnarray}
&&\hspace{-2mm}
I^{\mu\nu}(q,k_\perp)
~=~{N_c\over 64}\!\int\! {d\nu\over \pi\nu}{\sinh\pi\nu\over (1+\nu^2)\cosh^2\pi\nu}
\Big({k_\perp^2\over Q^2}\Big)^{\half-i\nu}
\nonumber\\
&&\hspace{1mm}
\times~\Big\{
\Big({9\over 4}+\nu^2\Big)\Big[1+{3\alpha_s\over 4\pi}c_F+{\alpha_sN_c\over 2\pi}\calf_1(\nu)\Big]P_1^{\mu\nu}
\nonumber\\
&&\hspace{1mm}
+~\Big({11\over 4}+3\nu^2\Big)\Big[1+{3\alpha_s\over 4\pi}c_F+{\alpha_sN_c\over 2\pi}\calf_2(\nu)\Big]P_2^{\mu\nu}
\nonumber\\
&&\hspace{1mm}
+~
\Big({1\over 8}+{\nu^2\over 2}\Big)
\Big[1+{3\alpha_s\over 4\pi}c_F+{\alpha_sN_c\over 2\pi}\calf_3(\nu)\Big]P_3^{\mu\nu}\Big\}
\label{nloifmom}
\end{eqnarray}
where (as usual, $\gamma\equiv \half+i\nu$)
\begin{eqnarray}
&&\hspace{-0mm}
\calf_{1(2)}(\nu)~=~\Phi_{1(2)}(\nu)+\chi_\gamma\Psi(\nu),
\label{calfs}\\
&&\hspace{0mm}
\calf_3(\nu)~=~F_6(\nu)+\Big(\chi_\gamma-{1\over\bamma\gamma}\Big)\Psi(\nu),
\nonumber\\
&&\hspace{0mm}
\Psi(\nu)~\equiv~\psi(\bamma)+2\psi(2-\gamma)-2\psi(4-2\gamma)-\psi(2+\gamma)
\nonumber
\end{eqnarray}
The structures $P_1$ and $P_2$ correspond to unpolarized structure functions $F_1(x_B)$ and $F_2(x_B)$. The third term vanishes for
nucleon structure function but contributes to polarized structure functions of a vector meson (or photon).

It is instructive to compare Eq. (\ref{nloifmom}) with the well-known double-integral representation 
of the leading-order impact factor (see e.g. Ref. \cite{loif}, \cite{mobzor}):
\begin{eqnarray}
&&\hspace{-0mm}
I^{\mu\nu}(q,k)~
=~{N_c\over 8\pi^2}\!\int_0^1\! dudv{k_\perp^2\over Q^2\baru u+k_\perp^2\barv v}
\Big\{(1-2\baru u
\nonumber\\
&&\hspace{0mm}
-~2\barv v+4\baru u\barv v)P_1^{\mu\nu}
+(1-2\baru u-2\barv v+12\baru u\barv v)P_2^{\mu\nu}
\nonumber\\
&&\hspace{22mm}
+~{2\baru u\barv v\over k_\perp^2}\big(\tildeP^{\mu\nu}\bark^2+\barP^{\mu\nu}\tildek^2\big)\Big\}
\label{knownloif}
\end{eqnarray}
(Note that the definition of $I^{\mu\nu}(q,k)$ differs in sign and $N_c$ from that of Ref. \cite{mobzor}).
It is easy to see that Eq. (\ref{knownloif}) is equal to LO terms in the r.h.s. of Eq. (\ref{nloifmom}).

\section{NLO BFKL for color dipoles}
For completeness, in this Section we present (linearized)  evolution equation for composite color dipoles
and discuss how it is related to usual NLO BFKL approach \cite{nlobfkl}. 
The evolution equation for forward matrix elements of color dipoles $\calu(z_{12})\equiv\langle\hat{\cal U}(z_1,z_2)\rangle$ reads \cite{nlobk,nlobksym}
\begin{eqnarray}
&&\hspace{-0mm}
2a{d\over da}{\cal U}_a(z)~=~{\alpha_sN_c\over 2\pi^2}
\!\int\!d^2z'~{z^2\over {z'}^2(z-z')^2}
\Big\{1
\nonumber\\
&&\hspace{-0mm} 
+~{\alpha_s\over 4\pi}\Big[b\big(\ln {z^2\mu^2\over 4} +2C\big)
-b{(z-z')^2-{z'}^2\over z^2}\ln{(z-z')^2\over {z'}^2}
\nonumber\\
&&\hspace{-0mm} 
+~
\big({67\over 9}-{\pi^2\over 3}\big)N_c-{10n_f\over 9}\Big]\Big\}
\Big[{\cal U}_a(z')+\calu_a(z-z')-{\cal U}_a(z)\Big]
\nonumber\\
&&\hspace{-0mm} 
+~{\alpha_s^2N_c^2\over 4\pi^3}\!\int\! d^2z'~{z^2\over {z'}^2}
\Big[-{1\over (z-z')^2}\ln^2{z^2\over {z'}^2}
\nonumber\\
&&\hspace{-0mm} 
+~F(z,z')+\Phi(z,z')\Big]~{\cal U}_a(z') +3{\alpha_s^2N_c^2\over 2\pi^2}\zeta(3){\cal U}_a(z)
\label{nlobfklu}
\end{eqnarray}
where 
\begin{eqnarray}
&&\hspace{-0mm}
F(z,z')~=~\Big(1+{n_f\over N_c^3}\Big){3(z,z')^2-2z^2{z'}^2\over16z^2{z'}^2}
\Big({2\over z^2}+{2\over {z'}^2}
\nonumber\\
&&\hspace{-0mm}
+~{z^2-{z'}^2\over z^2{z'}^2}\ln{z^2\over {z'}^2}\Big)
-~\Big[3+\Big(1+{n_f\over N_c^3}\Big)\Big(1-{(z^2+{z'}^2)^2\over 8z^2{z'}^2}
\nonumber\\
&&\hspace{-0mm}
+~{3z^4+3{z'}^4-2z^2{z'}^2\over 16z^4{z'}^4}(z,z')^2\Big)\Big]\!\int_0^\infty\!{dt\over z^2+t^2{z'}^2}\ln{1+t\over |1-t|} 
\nonumber\\
\label{F}
\end{eqnarray}
and 
\begin{eqnarray}
&&\hspace{-0mm}
\Phi(z,z')~=~{(z^2-{z'}^2)\over (z-z')^2(z+z')^2}
\Big[\ln{z^2\over {z'}^2}\ln{z^2{z'}^2(z-z')^4\over (z^2+{z'}^2)^4}
\nonumber\\
&&\hspace{-0mm}
+~2{\rm Li_2}\Big(-{{z'}^2\over z^2}\Big)-2{\rm Li_2}\Big(-{z^2\over {z'}^2}\Big)\Big]-~\Big(1
\label{Phi}\\
&&\hspace{-0mm}
-~{(z^2-{z'}^2)^2\over (z-z')^2(z+z')^2}\Big)\Big[\!\int_0^1-\int_1^\infty\Big]
{du\over (z-z'u)^2}\ln{u^2{z'}^2\over z^2}
\nonumber
\end{eqnarray}
Note that the kernel is a sum of the ``running-coupling'' part proportional to $b={11\over 3}N_c-{2\over 3}n_f$ and the conformal part,
see the discussion in Ref. \cite{nlolecture}. Here $\alpha_s\equiv\alpha_s(\mu)$ and $\mu$ is the normalization point in the $\overline{MS}$ scheme.

With the $k_T$-factorization in view let us  rewrite the evolution equation (\ref{nlobfklu}) in terms of   
\begin{equation}
\hspace{-0mm}
\calv_{a_m}(z_\perp)~\equiv~-\partial_\perp^2 \langle\!\langle\halu_{a_m}(z_\perp,0) \rangle\!\rangle
\label{calv}
\end{equation}
proportional to the dipole unintegrated gluon distribution $\cald(x_B, z_\perp,\mu)$
\begin{eqnarray}
&&\hspace{-0mm}
\calv_{x_B}(z_\perp,\mu)~=~{4\pi^2 x_B\over N_c}\alpha_s(\mu)\cald(x_B, z_\perp,\mu)
\label{dgTMDa}
\end{eqnarray}
where
\begin{eqnarray}
&&\hspace{-0mm}
\cald(x_B,z_\perp,\mu)~
\nonumber\\
&&\hspace{-0mm}\equiv~\big({2\over s}\big)^2\!\int\!{dz_\ast\over \pi x_B}
\langle p|{\rm Tr}\big\{\big[\infty p_1+z_\perp,{2\over s}z_\ast p_1+z_\perp\big]
\nonumber\\
&&\hspace{-0mm}
\hat{F}_{\bu\xi}\big({2\over s}z_\ast p_1+z_\perp\big)\big[{2\over s}x_\ast p_1+z_\perp, -\infty p_1+z_\perp]
\nonumber\\
&&\hspace{-0mm}
\times~[-\infty p_1,0]\hat{F}_\bu^{~\xi}(0)[0,\infty p_1]\big\}|p\rangle^{a_m=x_B}
\label{dgTMD}
\end{eqnarray}
Hereafter we use the notation
$$
[x,y]~\equiv~{\rm Pexp}\Big\{ig\!\int_0^1\! du ~(x-y)^\mu A_\mu(ux+\baru y)\Big\}
$$
for the gauge link connecting points $x$ and $y$.
The color dipole is renorm-invariant  so $\cald$  depends on $\mu$ to compensate $g^2(\mu)$ dependence.

The Fourier transform  
$$
\cald(x_B,k_\perp)=\int\! d^2z\, e^{i(k,z)_\perp}\cald(x_B,z_\perp)
$$ 
is called the dipole gluon TMD (transverse momentum dependent distribution).
Note, however, that the dipole gluon TMD defined above differs from the definition 
\begin{eqnarray}
&&\hspace{-0mm}
\tilde{\cald}(x_B,z_\perp,\mu,\eta)~
\nonumber\\
&&\hspace{-0mm}=~\big({2\over s}\big)^2\!\int\!{dz_\ast\over \pi x_B}e^{-ix_Bz_\ast}
\langle p|{\rm Tr}\big\{\big[\infty p_1+z_\perp,{2\over s}z_\ast p_1+z_\perp\big]
\nonumber\\
&&\hspace{-0mm}
\hat{F}_{\bu\xi}\big({2\over s}z_\ast p_1+z_\perp\big)\big[{2\over s}x_\ast p_1+z_\perp, -\infty p_1+z_\perp]
\nonumber\\
&&\hspace{-0mm}
[-\infty p_1,0]\hat{F}_\bu^{~\xi}(0)[0,\infty p_1]\big\}|p\rangle^{\eta}
\label{gTMD}
\end{eqnarray}
which reduces to the usual parton density at $z_\perp=0$. It should be emphasized that Eq.
(\ref{gTMD}) is a more complex operator than (\ref{dgTMD}). The difference is especially clear in 
the case of ${\cal N}=4$ theory: the dipole gluon TMD (\ref{dgTMD}) is UV finite while Eq. (\ref{gTMD})
is UV divergent so it needs additional UV counterterms, see the discussion in \cite{collinsbook}.
(These UV divergent terms  are directly proportional to $x_B$ so they vanish for the definition (\ref{dgTMD})). 
Also, the role of parameter $x_B$ is different in the two definitions: in Eq. (\ref{dgTMD}) 
it is defined as a rapidity cutoff $a_m$ while in Eq. (\ref{gTMD}) the rapidity cutoff $\eta$ should be imposed separately from $x_B$. 
 
Differentiating Eq. (\ref{nlobfklu}) two times with respect to $z$ we obtain
the NLO BFKL evolution for dipole gluon TMD (\ref{dgTMD}) in the form\\
\begin{eqnarray}
&&\hspace{-0mm}
2a{d\over da}\calv_a(z)~
\nonumber\\
&&\hspace{-0mm} 
=~{\alpha_sN_c\over \pi^2}\!\int\!d^2z'
\Big\{\Big(1+~{\alpha_sb\over 4\pi}\Big[\ln {z^2\mu^2\over 4} +2C+{67N_c\over 9b}
\nonumber\\
&&\hspace{-0mm} 
-~{\pi^2N_c\over 3b}-{10n_f\over 9b}\Big]\Big)
\big[{\calv_a(z')\over (z-z')^2}-{(zz')\calv_a(z)\over z'^2(z-z')^2}\big]
\nonumber\\
&&\hspace{-0mm} 
+~{\alpha_sb\over2\pi}~{\calv_a(z')-\calv_a(z)\over (z-z')^2}\ln{(z-z')^2\over{z'}^2}
\nonumber\\
&&\hspace{-0mm} 
+~{\alpha_sN_c\over 4\pi}
\Big[-{\ln^2(z^2/{z'}^2)\over (z-z')^2}
+~F(z,z')+\Phi(z,z')\Big]\calv_a(z') \Big\}
\nonumber\\
&&\hspace{-0mm} 
+~3{\alpha_s^2N_c^2\over 2\pi^2}\zeta(3)\calv_a(z)
\label{nlobfkld}
\end{eqnarray}

Next we need to perform the Fourier transformation of Eq. (\ref{nlobfkld}). It can be demonstrated that
\begin{eqnarray}
&&\hspace{-5mm}
\int\! {d^2qd^2q'\over 4\pi^2}~e^{i(q,z)-i(q',z')}\Big[-{\ln^2(q^2/{q'}^2)\over (q-q')^2}+F(q,q')
\nonumber\\
&&\hspace{-5mm}
+~\Phi(q,q')\big]~=~-{\ln^2(z^2/{z'}^2)\over (z-z')^2}+F(z,z')+\Phi(z,z')
\label{furye}
\end{eqnarray}
so the conformal part of the kernel looks the same in coordinate and momentum representations.

Performing also the Fourier transformation of the running-coupling part one obtains 
the momentum-representation kernel in the form\\
\begin{eqnarray}
&&\hspace{-0mm}
2a{d\over da}{\cal V}_a(k)~=~{\alpha_sN_c\over \pi^2}
\!\int\!d^2k'
\Big\{\Big[{\calv_a(k')\over (k-k')^2}-{(k,k')\calv_a(k)\over {k'}^2(k-k')^2}\Big]
\nonumber\\
&&\hspace{-0mm} 
\times~\Big(1+~{\alpha_sb\over 4\pi}\Big[\ln{\mu^2\over k^2}+{N_c\over b}\big({67\over 9}-{\pi^2\over 3}
-~{10n_f\over 9N_c}\big)\Big]\Big)-~{b\alpha_s\over 4\pi}
\nonumber\\
&&\hspace{-0mm} 
\times~\Big[{\calv_a(k')\over (k-k')^2}\ln{(k-k')^2\over {k'}^2}-{k^2\calv_a(k)\over{k'}^2(k-k')^2}\ln{(k-k')^2\over k^2}\Big]
\nonumber\\
&&\hspace{-0mm} 
+~{\alpha_sN_c\over 4\pi}
\Big[-{\ln^2(k^2/{k'}^2)\over (k-k')^2}
+~F(k,k')+\Phi(k,k')\Big]~{\cal V}_a(k') \Big\}
\nonumber\\
&&\hspace{-0mm} 
+~3{\alpha_s^2N_c^2\over 2\pi^2}\zeta(3){\cal V}_a(k)
\label{nlobfklkernel}
\end{eqnarray}
where $\calv(k)\equiv\int\! dz~ e^{-i(k,z)_\perp}\calv(z)$. 

In terms of Mellin projections ($\gamma\equiv\half+i\nu$ as usual)
\begin{eqnarray}
&&\hspace{-0mm}
\calv(k)~=~\sum_{n=0}^\infty\!\int{d\nu\over 2\pi^2}\calv(n,\nu)(k^2)^{\gamma-1}\big({\tildek/\bark}\big)^{n/2}
\nonumber\\
&&\hspace{-0mm} 
\calv(n,\nu)~=~\int\! d^2k~(k^2)^{-\gamma}\big({\bark/\tildek}\big)^{n/2}\calv(k)
\nonumber\\
&&\hspace{-0mm} 
=~4\pi i^n{\Gamma\big(\bamma+{n\over 2}\big)\over\Gamma\big(\gamma+{n\over 2}\big)}\!\int\! d^2z(z^2)^{\gamma-1}\big({\barz\over\tildez}\big)^{n/2}\calv(z)
\end{eqnarray}
the kernel (\ref{nlobfklkernel}) takes the form\\
\begin{eqnarray}
&&\hspace{-0mm}
2a{d\over da}{\cal V}_a(n,\nu)~
\label{evolva}\\
&&\hspace{-0mm} 
=~{\alpha_sN_c\over \pi}
\Big\{\chi(n,\gamma)
+~{\alpha_sN_c\over 4\pi}\Big[{b\over N_c}\chi(n,\gamma)\Big(\ln\mu^2
+ {d\over d\gamma}
\nonumber\\
&&\hspace{-0mm} 
-~{\chi(n,\gamma)\over 2}+{\chi'(n,\gamma)\over 2\chi(n,\gamma)} \Big)
+f(n,\gamma)\Big]\Big\}{\cal V}_a(n,\nu)
\nonumber
\end{eqnarray}
Here
\begin{eqnarray}
&&\hspace{-0mm} 
f(n,\gamma)~=~\big[{67\over 9}-{\pi^2\over 3}-{10n_f\over 9N_c}\big]\chi(n,\gamma)
-~\chi''(n,\gamma)
\nonumber\\
&&\hspace{-0mm} 
+~F(n,\gamma)-2\Phi(n,\gamma)-2\Phi(n,1-\gamma)+6\zeta(3)
\end{eqnarray}
where
\begin{eqnarray}
&&\hspace{-1mm}
F(n,\gamma)~
=~\Big\{-\Big[3+\Big(1+{n_f\over N_c^3}\Big)
{2+3\gamma\bar{\gamma}\over (3-2\gamma)(1+2\gamma)}\Big]\delta_{0n}
\nonumber\\
&&\hspace{-1mm}
+~\Big(1+{n_f\over N_c^3}\Big)
{\gamma\bar{\gamma}\over 2(3-2\gamma)(1+2\gamma)}\delta_{2n}\Big\}
{\pi^2\cos\pi\gamma\over(1-2\gamma)\sin^2\pi\gamma}
\nonumber
\end{eqnarray}
and
\begin{eqnarray}
&&\hspace{-1mm}
\Phi(n,\gamma)~=~\int_0^1\!{dt\over 1+t}~t^{\gamma-1+{n\over 2}}
\Big\{{\pi^2\over 12}-{1\over 2}\psi'\Big({n+1\over 2}\Big)
\nonumber\\
&&\hspace{-1mm}
-{\rm Li}_2(t)-{\rm Li}_2(-t)
-~\Big[\psi(n+1)-\psi(1)+\ln(1+t)
\nonumber\\
&&\hspace{-1mm}+\sum_{k=1}^\infty{(-t)^k\over k+n}\Big]\ln t
-\sum_{k=1}^\infty{t^k\over (k+n)^2}[1-(-1)^k]\Big\}
\label{fin}
\end{eqnarray}

To compare to NLO BFKL from Ref. \cite{nlobfkl} one should rewrite above equation
in terms of
\begin{equation}
\call(k)~=~{1\over g^2(k)}\calv(k)
\end{equation}
since two gluons in the dipole $\calu(k)$ come with extra $g^2$ factor.  The Eq. (\ref{evolva}) turns into\\
\begin{eqnarray}
&&\hspace{-0mm}
2a{d\over da}{\cal L}_a(k)~=~{\alpha_s(k^2)N_c\over \pi^2}
\!\int\!d^2k'
\Big\{\Big[{\call_a(k')\over (k-k')^2}
\nonumber\\
&&\hspace{-0mm} 
-~{(k,k')\call_a(k)\over {k'}^2(k-k')^2}\Big]\Big[1+~{\alpha_sN_c\over 4\pi}\Big({67\over 9}-{\pi^2\over 3}
-~{10n_f\over 9N_c}\Big)\Big]-
\nonumber\\
&&\hspace{-0mm} 
-~{b\alpha_s\over 4\pi}\Big[{\call_a(k')\over (k-k')^2}-{k^2\call_a(k)\over{k'}^2(k-k')^2}\Big]\ln{(k-k')^2\over k^2}
\nonumber\\
&&\hspace{-0mm} 
+~{\alpha_sN_c\over 4\pi}
\Big[-{\ln^2(k^2/{k'}^2)\over (k-k')^2}
+~F(k,k')+\Phi(k,k')\Big]~{\cal L}_a(k') \Big\}
\nonumber\\
&&\hspace{-0mm} 
+~3{\alpha_s^2N_c^2\over 2\pi^2}\zeta(3){\cal L}_a(k)
\label{nlobfklkerneL}
\end{eqnarray}
with the eigenvalues\\
\begin{eqnarray}
&&\hspace{-0mm} {\alpha_s(k^2)N_c\over \pi^2}
\!\int\!d^2k'
\Big\{\Big[{({k'}^2/k^2)^{-\bamma}e^{in\phi}\over (k-k')^2}-{(k,k')\over {k'}^2(k-k')^2}\Big]
\nonumber\\
&&\hspace{-0mm} 
\times~\Big[1+~{\alpha_sN_c\over 4\pi}\big({67\over 9}-{\pi^2\over 3}-{10n_f\over 9N_c}\big)\Big]
\nonumber\\
&&\hspace{-0mm} 
-~{b\alpha_s\over 4\pi}\Big[{({k'}^2/k^2)^{-\bamma}e^{in\phi}\over (k-k')^2}-{k^2\over{k'}^2(k-k')^2}\Big]\ln{(k-k')^2\over k^2}
\nonumber\\
&&\hspace{-0mm} 
+~{\alpha_sN_c\over 4\pi}
\Big[-{\ln^2(k^2/{k'}^2)\over (k-k')^2}
+~F(k,k')+\Phi(k,k')\Big]
\nonumber\\
&&\hspace{-0mm} 
\times~({k'}^2/k^2)^{-\bamma}e^{in\phi} \Big\}+~3{\alpha_s^2N_c^2\over 2\pi^2}\zeta(3)
\nonumber\\
&&\hspace{-0mm}
=~{\alpha_s(k^2)N_c\over \pi}
\Big\{\chi(n,\gamma)
\nonumber\\
&&\hspace{-0mm} 
+~{\alpha_sN_c\over 4\pi}\Big[
-{b\over 2N_c}[\chi^2(n,\gamma)+\chi'(n,\gamma)] 
\label{nlobfkleigenvalues}
+f(n,\gamma)\Big]\Big\}\end{eqnarray}
which coincide with eigenvalues of the kernel \cite{nlobfkl} of  the partial wave of the forward reggeized gluon scattering amplitude
\begin{equation}
\omega G_\omega(q,q')=\delta^{(2)}(q-q')+\int\! d^2p~ K(q,p)G_\omega(p,q') 
\label{wgw}
\end{equation}
\begin{eqnarray}
&&\hspace{-1mm}
\int\! d^2p \Big({p^2\over q^2}\Big)^{\gamma-1}e^{in\phi}K(q,p)~=~
{\alpha_s(q)\over \pi}N_c\Big[\chi(n,\gamma)
\nonumber\\
&&\hspace{-0mm}
+~{\alpha_sN_c\over 4\pi}\Big(f(n,\gamma)-{b\over 2N_c}[\chi'(n,\gamma)+\chi^2(n,\gamma)]\Big)
\Big]
\label{lip1}
\end{eqnarray}
This is somewhat surprising since the evolution of the composite (in ${\cal N}=4$ SYM - conformal) dipole
with respect to $a$ gives the evolution of  forward reggeized gluon scattering amplitude with respect to rapidity $\eta$ 
(of which $\omega$ is the Mellin transform). To illustrate the transition between the two evolutions let us consider 
the calculation of the dipole evolution directly from the NLO BFKL for reggeized gluons.

The  impact factor $\Phi_A(q)$ for the color dipole ${\cal U}(x,y)$ is proportional to
$\alpha_s(q)(e^{iqx}-e^{iqy})(e^{-iqx}-e^{-iqy})$ so one obtains the cross section of the scattering of color dipole in the form
\begin{eqnarray}
&&\hspace{-1mm}
\calu^\eta(x)~=~\!\int\!{d^2 q\over q^2}
{d^2 q'\over {q'}^2}{\alpha_s(q)\over 4\pi^2}
(e^{iqx}-1)(e^{-iqx}-1)\Phi_B(q')
\nonumber\\
&&\hspace{-0mm}
\times~
\!\int_{a-i\infty}^{a+i\infty}\!{d\omega\over 2\pi i}\Big({se^\eta \over qq'}\Big)^\omega
G_\omega(q,q')
\label{colordip}
\end{eqnarray}
 $\Phi_B(q')$ is the target impact factor. 
 To get the evolution equation with respect to rapidity one should change the energy scale to ${q'}^2$. 
\begin{eqnarray}
&&\hspace{-1mm}
\calu^\eta(x)~=~\!\int\!{d^2 q\over q^2}
{d^2 q'\over {q'}^2}{\alpha_s(q)\over 4\pi^2}
(e^{iqx}-1)(e^{-iqx}-1)\Phi_B(q')
\nonumber\\
&&\hspace{-0mm}
\times~
\!\int_{a-i\infty}^{a+i\infty}\!{d\omega\over 2\pi i}\Big({se^\eta \over {q'}^2}\Big)^\omega
\tilde{G}_\omega(q,q')
\label{lip2}
\end{eqnarray}
where $\tilde{G}_\omega(q,q')$ is the modified kernel with eigenvalues shifted by 
$2\chi(n,\nu)\chi'(n,\nu){\alpha_s^2N_c^2\over 4\pi^2}$, see Ref. \cite{nlobfkl}.
 The corresponding equation for 
$\calv^\eta(n,\nu)$ takes the form \cite{nlobk}
\begin{eqnarray}
&&\hspace{-0mm}
{d\over d\eta}{\cal V}^\eta(n,\nu)~=~{\alpha_sN_c\over \pi}
\Big\{\chi(n,\gamma)+~{\alpha_sN_c\over 4\pi}\Big[{b\over N_c}\chi(n,\gamma)
\nonumber\\
&&\hspace{-0mm} 
\times~
\Big(\ln\mu^2
+ {d\over d\gamma}
-{\chi(n,\gamma)\over 2}+{\chi'(n,\gamma)\over 2\chi(n,\gamma)} \Big)
\nonumber\\
&&\hspace{0mm}
+~f(n,\gamma)+2\chi(n,\nu)\chi'(n,\nu)\Big]\Big\}{\cal V}^\eta(n,\nu)
\label{evolveta}
\end{eqnarray}
Let us demonstrate that it agrees with Eq. (\ref{evolva}). The Mellin projection of composite dipole can be obtained from Eq. (\ref{confodipola}): 
\begin{eqnarray}
&&\hspace{-0mm}
\calv^a(n,\nu)~
=~\Big(1+{\alpha_sN_c\over 4\pi}\Big[\chi^2(n,\gamma)+3\chi'(n,\gamma)
\nonumber\\
&&\hspace{20mm} +2\chi(n,\gamma)\big(\ln a-2\eta+2C-{d\over d\gamma}\big)\Big]
\nonumber\\
&&\hspace{0mm} 
+~\big({\alpha_sN_c\over 4\pi}\big)^2\Big\{2\chi^2(n,\gamma)(\ln a-2\eta-{d\over d\gamma})^2
\nonumber\\
&&\hspace{0mm}
+~2(\ln a-2\eta)\Big[{b\over N_c}\chi(n,\gamma)
\Big(\ln\mu^2
+ {d\over d\gamma}-{\chi(n,\gamma)\over 2}
\nonumber\\
&&\hspace{0mm}
+~{\chi'(n,\gamma)\over 2\chi(n,\gamma)} \Big)
+~f(n,\gamma)+3\chi(n,\gamma)\chi'(n,\gamma)
\nonumber\\
&&\hspace{0mm}
+\chi^3(n,\gamma)\Big]+X(n,\nu)\Big\}\calv^\eta(n,\nu)
\label{relation}
\end{eqnarray}
The first-order term can be derived from Eq. (\ref{confodipola}) while the term $\sim\alpha_s^2$ can be restored
from the condition ${d\over d\eta}\calv^a(n,\nu)=0$ up to an unknown function $X(n,\nu)$ which requires NNLO calculation
(and does not contribute to the NLO evolution).
Now one can see that the derivative with respect to $a$ gives Eq. (\ref{evolva}).  Thus, the transition between evolution of the composite dipole 
$\calv^a(n,\nu)$ with respect to $a$ and the rapidity evolution of the dipole $\calv^\eta(n,\nu)$ corresponds to the shift in eigenvalues 
on the function $2\chi(n,\nu)\chi'(n,\nu){\alpha_s^2N_c^2\over 4\pi^2}$ - the same transition that describes the shift of eigenvalues
when going from energy scale $qq'$ to ${q'}^2$ in formula (\ref{lip2}).

\section{Conclusions}
Let us present again the $k_T$ factorization formula for DIS in the next-to-leading order:
\begin{eqnarray}
&&\hspace{-0mm}
\int\! d^4x ~e^{iqx} \langle p|T\{\hatj_{\mu}(x)\hatj_\nu(0)\}|p\rangle~
\nonumber\\
&&\hspace{-0mm}
=~{s\over 2}\big(\sum e_i^2\big)\int\! {d^2k_\perp\over 4\pi^2k_\perp^2}~ I_{\mu\nu}(q,k_\perp) \calv_{a_m=x_B}(k_\perp)
\label{ktfacv}
\end{eqnarray}
where $I_{\mu\nu}(q,k_\perp)$ is given by Eq. (\ref{nloifmom}) and the evolution equation for $\calv_a(k_\perp)$ by Eq. (\ref{nlobfklkernel}) (or Eq. (\ref{evolva}) in the
Mellin representation).

The analytic NLO photon impact factor 
in momentum space for the pomeron contribution (\ref{nloifmom}) and 
the NLO $k_{\rm T}$ factorization formula (\ref{ktfacv}) for the deep inelastic scattering
are the main results of this paper.  
 
 Since the composite dipole (68) obeys the same equation as forward scattering amplitude of two reggeized gluons (71) the impact factor (\ref{nloifmom}) may be obtained as an NLO amplitude of
 emission of two reggeized gluons by the virtual photon. There were several attempts in the literature to obtain this amplitude 
 \cite{attempts}, but at present  such impact factor
is known only as a combination of analytical and numerical results \cite{bart1}.  Indeed, 
in Ref. \cite{attemptNLO} the $\gamma^*-\gamma^*$ cross-section 
has been calculated using only the LO impact-factor and the LO and NLO BFKL amplitude for two reggized gluons. The authors
explain in the paper that the NLO impact-factor as know at present in Ref. \cite{bart1}
is difficult to handle for numerical calculation since it is not in a full analytic form. 
On the other hand, the result of this paper, provided that one knows the solution of the NLO BFKL with the running
coupling constant, allows to compute the full NLO total cross section for the 
$\gamma^\ast-\gamma^\ast$ scattering process.
 
An attempt to calculate the NLO impact factor in an analytic form using an approach based on the 
analytic properties of the amplitude can be found in Ref. \cite{attemFadin}.

In the past few years there has been some activity on the calculation of NLO impact factor of other processes as well:
in Ref. \cite{CapIvaMurdPapa} the calculation of the NLO impact factor for Mueller-Navelet jets has been performed, while the impact factor for 
virtual photon to light vector meson transition has been performed in Ref. \cite{IvaKotyPapa}.

It would be also instructive to compare our result (\ref{opeconfin}) for the coefficient in front of the four-Wilson-line operator 
(relevant for the structure functions of DIS off a large nucleus) 
to similar result for the NLO impact factor obtained recently in Ref. \cite{Beuf:2011xd} using the dipole model. 
 However, as we already mentioned
 our final NLO result (\ref{nloifmom}) is defined as a coefficient function in front of composite operator 
(\ref{confodipola}) defined with a counterterm which restores the 
 conformal invariance in ${\cal N}=4$ amplitudes and in our case leads to the conformal impact factor 
 (since the impact factor is given by tree diagrams
 it should be conformally invariant even in QCD). As a consequence, the impact 
factor depends on a new parameter $a$ (an analog of the factorization scale $\mu$ 
in  usual OPE) which we chose in such a way that 
all the energy dependence is shifted in to the matrix element, leaving the impact factor energy-scale invariant.
To compare with the result of Ref. \cite{Beuf:2011xd} representing the coefficient function of a usual dipole 
(without counterterm subtraction) we should trace one step back and look at the impact factor $I^{\rm NLO}_{\mu\nu}(z_1,z_2,z_3;\eta)$ given by (\ref{nloify}).
One should then perform Fourier transformation to momentum space 
with respect to the positions $x$ and $y$ of the two electromagnetic currents  in formula (\ref{opeq}) 
and compare it to the result (58) from Ref. \cite{Beuf:2011xd} integrated over $z_1$ (and $z_2$ when appropriate). 
Hopefully, after these integrations the two results will coincide. 

This work was supported by contract
 DE-AC05-06OR23177 under which the Jefferson Science Associates, LLC operate the Thomas Jefferson National Accelerator Facility, 
and by the grant DE-AC02-05CH11231.

 \section{Appendix A}
 There is a subtle point in the Fourier transformation of the Eq. (\ref{opemellin}) in the forward case (cf. Ref. \cite{lip86}). To illustrate it, consider the simplest term in the r.h.s. of Eq. (\ref{nloifresult})  
\begin{eqnarray}
&&\hspace{-1mm}
-{\partial\kappa^\alpha\over\partial x^\mu}{\partial\kappa^\beta\over\partial y^\nu}{2\over\kappa^2}
\Big(g^{\alpha\beta}-2{\kappa^\alpha\kappa^\beta\over\kappa^2}\Big)\calr^2~
\nonumber\\
&&\hspace{0mm}
=~{2\calr^2\over (x-y)^2}\Big(g_{\mu\nu}-2{(x-y)_\mu(x-y)_\nu\over (x-y)^2}\Big)
\end{eqnarray}
The corresponding contribution to the T-product of currents (\ref{ifresult}) is proportional to
\begin{equation}
\hspace{-0mm}
\!\int\! {dz_1 dz_2\over z_{12}^4}~\halu(z_1,z_2){\calr^2\over (x-y)^6}\Big[g_{\mu\nu}-2{(x-y)_\mu(x-y)_\nu\over (x-y)^2}\Big]~
\end{equation}
Consider now the Fourier transform of this equation for the case of forward scattering
\begin{eqnarray}
&&\hspace{-0mm}
\int\! d^4 x d^4y ~\delta(y_\bu)~e^{iq\cdot (x-y)}
\!\int\! {dz_1 dz_2\over z_{12}^4}{\calr^2\over (x-y)^6}
\label{fla53}\\
&&\hspace{0mm}
~\Big[g_{\mu\nu}-2{(x-y)_\mu(x-y)_\nu\over (x-y)^2}\Big]
\calu(z_{12})
\nonumber
\end{eqnarray}
where $\calu(z_{12})=\langle\hat{\cal U}(z_1,z_2)\rangle$. 
We have calculated such integrals by using the representation of the type 
\begin{eqnarray}
&&\hspace{-0mm}
\calr^2~
=~\int\!{d\nu\over\pi^2}~\nu^2r(\nu)\!\int\! d^2z_0
\Big({z_{12}^2\over z_{10}^2z_{20}^2}\Big)^{\gamma}\Big[{\kappa^2\over (2\kappa\cdot\zeta_0)^2}\Big]^{\bamma}
\nonumber
\end{eqnarray}
based on the decomposition (\ref{lobzor120}) of transverse $\delta$-functions. 
(For this example $r(\nu)~=~ B(\gamma,\gamma)\Gamma(1+\gamma)\Gamma(2-\gamma)$). 
The integral over $z_1$ and $z_2$ in Eq. (\ref{fla53}) is of the form
\begin{eqnarray}
&&\hspace{-0mm}
\int\! {dz_1 dz_2\over z_{12}^4}\Big({z_{12}^2\over z_{10}^2z_{20}^2}\Big)^{\half+i\nu}f(z_{12}^2)
\label{fla54}\\
&&\hspace{0mm}=~\int\!{d\mu\over 2\pi}
\!\int\! {dz_1 dz_2\over z_{12}^4}\Big({z_{12}^2\over z_{10}^2z_{20}^2}\Big)^{\half+i\nu}(z_{12}^2)^{\half-i\mu}f(\mu)
\nonumber
\end{eqnarray}
where $f(\mu)~=~\int\! dz_{12}^2 ~(z_{12}^2)^{-{3\over 2}+i\mu}f(z_{12}^2)$. To calculate the integral in the r.h.s. this equation, we take
the orhtogonality condition for conformal eigenfunctions \cite{lip86}
\begin{eqnarray}
&&\hspace{-0mm}
\int\! {d^2z_1 d^2z_2\over z_{12}^4}~\Big({z_{12}^2\over z_{10}^2z_{20}^2}\Big)^{\half +i\nu}\!
\Big({z_{12}^2\over z_1^2z_2^2}\Big)^{\half -i\mu}~=~\Big[\delta(\nu-\mu)
\nonumber\\
&&\hspace{0mm}
\times~\delta^{(2)}(z_0)~+~\delta(\nu+\mu)(z_0^2)^{-1+2i\mu}{2i\mu B(\half+i\mu)\over \pi B(\half-i\mu)}\Big]  {\pi^4\over 2\nu^2}   
\nonumber\\
\label{liportho}
 \end{eqnarray}
and perform the inversion $z_i\rightarrow {z_i\over z^2}$. We obtain
\begin{eqnarray}
&&\hspace{-0mm}  
\int\! {d^2z_1 d^2z_2\over z_{12}^4}~\Big({z_{12}^2\over z_{10}^2z_{20}^2}\Big)^{\half +i\nu}
(z_{12}^2)^{\half -i\mu}~=~ {\pi^4\over 2\nu^2}\Big[\delta(\nu-\mu)
\nonumber\\
&&\hspace{0mm}
\times~(z_0^2)^{-1-2i\nu}\delta^{(2)}\big({1\over z_0}\big)~
+~\delta(\nu+\mu){2i\mu B(\half+i\mu)\over \pi B(\half-i\mu)}\Big]  
\label{fla56}
 \end{eqnarray}
so Eq. (\ref{fla54}) turns into
\begin{eqnarray}
&&\hspace{-0mm}
\int\! {dz_1 dz_2\over z_{12}^4}\Big({z_{12}^2\over z_{10}^2z_{20}^2}\Big)^{\half+i\nu}f(z_{12}^2)
\label{fla57}\\
&&\hspace{0mm}
=~{\pi^3\over 4\nu^2}(z_0^2)^{-1-2i\nu}f(\nu)\delta^{(2)}\big({1\over z_0}\big)-{i\pi^2B(\half-i\nu)\over 2\nu B(\half+i\nu)}f(-\nu)
\nonumber
\end{eqnarray}
Substituting this equation with $f(z_{12}^2)=\calu(z_{12})=z_{12}^2\calv(z_{12})$ into Eq. (\ref{fla53}) we get
\begin{eqnarray}
&&\hspace{-2mm}
{\pi\over 4}\!\int\! d^4 x d^4y ~\delta(y_\bu)~{e^{iq\cdot (x-y)}\over (x-y)^6}\Big[g_{\mu\nu}-2{(x-y)_\mu(x-y)_\nu\over (x-y)^2}\Big]
\nonumber\\\
&&\hspace{-2mm}
\times~r(\nu)\Big\{\Big[{(x-y)^2x_\ast y_\ast\over (x_\ast-y_\ast)^2}\Big]^\bamma\calv(\nu)
\nonumber\\\
&&\hspace{-2mm}+~{B(\bamma)\over B(\gamma)}\Big[{(x-y)^2x_\ast y_\ast\over (x_\ast-y_\ast)^2}\Big]^\gamma\calv(-\nu)\Big\}
\end{eqnarray}
where $V(\nu)~\equiv~{1\over\pi}\!\int\! d^2z (z^2)^{-\half+i\nu}\calv(z)$. Here the first term comes from $z_0=\infty$ while the second
from finite $z_0$. It is easy to see that the two terms coincide after change of integration variable $\nu\leftrightarrow -\nu$ so effectively the contribution of the integral over finite $z_0$ is doubled:
\begin{eqnarray}
&&\hspace{-0mm}
\int\! d^4 x d^4y ~\delta(y_\bu)~e^{iq\cdot (x-y)}
\!\int\! {dz_1 dz_2\over z_{12}^4}{\calr^2\over (x-y)^6}
\nonumber\\
&&\hspace{0mm}
~\Big[g_{\mu\nu}-2{(x-y)_\mu(x-y)_\nu\over (x-y)^2}\Big]
\calu(z_{12})
\nonumber\\
&&\hspace{-0mm}
=~{\pi\over 2}\!\int\! d^4 x d^4y ~\delta(y_\bu)\Big[g_{\mu\nu}-2{(x-y)_\mu(x-y)_\nu\over (x-y)^2}\Big]
\nonumber\\\
&&\hspace{0mm}
\times~r(\nu){B(\bamma)\over B(\gamma)}{e^{iq\cdot (x-y)}\over (x-y)^6}\Big[{(x-y)^2x_\ast y_\ast\over (x_\ast-y_\ast)^2}\Big]^\gamma V(-\nu)
\end{eqnarray}
Since we have not used the explicit form of the Lorentz structure in $\mu$ and $\nu$
 indices, it is clear that the doubling effect is general for any contribution to forward Fourier transform of Eq. (\ref{opemoment}) (see also the discussion 
 of zero transfer momentum limit in Ref. \cite{lip86}).

 
\end{document}